\pdfoutput=1
\documentclass[10pt,letterpaper]{article}

\usepackage{times}
\usepackage{xspace}
\usepackage[dvipsnames,table]{xcolor}
\usepackage{graphicx}
\usepackage{amsmath}
\usepackage{amssymb}
\usepackage{bm}
\usepackage{booktabs}
\usepackage{makecell}
\usepackage[figuresright]{rotating}
\usepackage[numbers,sort&compress]{natbib}
\usepackage[format=plain,labelformat=simple,labelsep=period,font=small,compatibility=false]{caption}
\usepackage[font=footnotesize,skip=3pt,subrefformat=parens]{subcaption}
\usepackage[hyphens]{url}
\usepackage{wrapfig}

\usepackage[pagebackref,breaklinks,colorlinks,allcolors=blue]{hyperref}

\usepackage[capitalize]{cleveref}
\crefname{section}{Sec.}{Secs.}
\Crefname{section}{Section}{Sections}
\Crefname{table}{Table}{Tables}
\crefname{table}{Tab.}{Tabs.}

\makeatletter
\DeclareRobustCommand\onedot{\futurelet\@let@token\@onedot}
\def\@onedot{\ifx\@let@token.\else.\null\fi\xspace}

\makeatother

\newcommand\blfootnote[1]{%
  \begingroup
  \renewcommand\thefootnote{}\footnote{#1}%
  \addtocounter{footnote}{-1}%
  \endgroup
}

\makeatletter
\font\elvbf = ptmb scaled 1100
\font\tenbf = ptmb scaled 1000

\def\@maketitle{%
   \newpage
   \null
   \vskip .375in
   \begin{center}%
      {\Large \bf \@title \par}%
      \vspace*{24pt}{%
        \large
        \lineskip .5em
        \begin{tabular}[t]{c}%
          \@author
        \end{tabular}\par}%
      \vskip .5em
      \vspace*{12pt}%
   \end{center}%
}

\def\abstract{%
   \thispagestyle{empty}%
   \centerline{\large\bf Abstract}%
   \vspace*{12pt}\noindent%
   \it\ignorespaces%
}

\def\customsection{\@startsection {section}{1}{\z@}
   {-10pt plus -2pt minus -2pt}{7pt} {\large\bf}}
\def\customssect#1{\customsection*{#1}}
\def\customsect#1{\customsection{\texorpdfstring{\hskip -1em.~}{}#1}}
\def\section{\@ifstar\customssect\customsect}

\def\customsubsection{\@startsection {subsection}{2}{\z@}
   {-8pt plus -2pt minus -2pt}{5pt} {\elvbf}}
\def\customssubsect#1{\customsubsection*{#1}}
\def\customsubsect#1{\customsubsection{\texorpdfstring{\hskip -1em.~}{}#1}}
\def\subsection{\@ifstar\customssubsect\customsubsect}

\def\customsubsubsection{\@startsection {subsubsection}{3}{\z@}
   {-6pt plus -2pt minus -2pt}{3pt} {\tenbf}}
\def\customssubsubsect#1{\customsubsubsection*{#1}}
\def\customsubsubsect#1{\customsubsubsection{\texorpdfstring{\hskip -1em.~}{}#1}}
\def\subsubsection{\@ifstar\customssubsubsect\customsubsubsect}
\makeatother

\title{\textit{VoroLight}: Learning Voronoi Surface Meshes via Sphere Intersection}

\author{
Jiayin Lu$^{1*}$ \quad 
Ying Jiang$^{1*}$ \quad 
Yumeng He$^{2*}$ \quad 
Yin Yang$^{3}$ \quad
Chenfanfu Jiang$^{1}$
}

\begin{document}

\setcounter{footnote}{0}

\maketitle

\vspace{-1em}
\begin{center}
  \includegraphics[width=\textwidth]{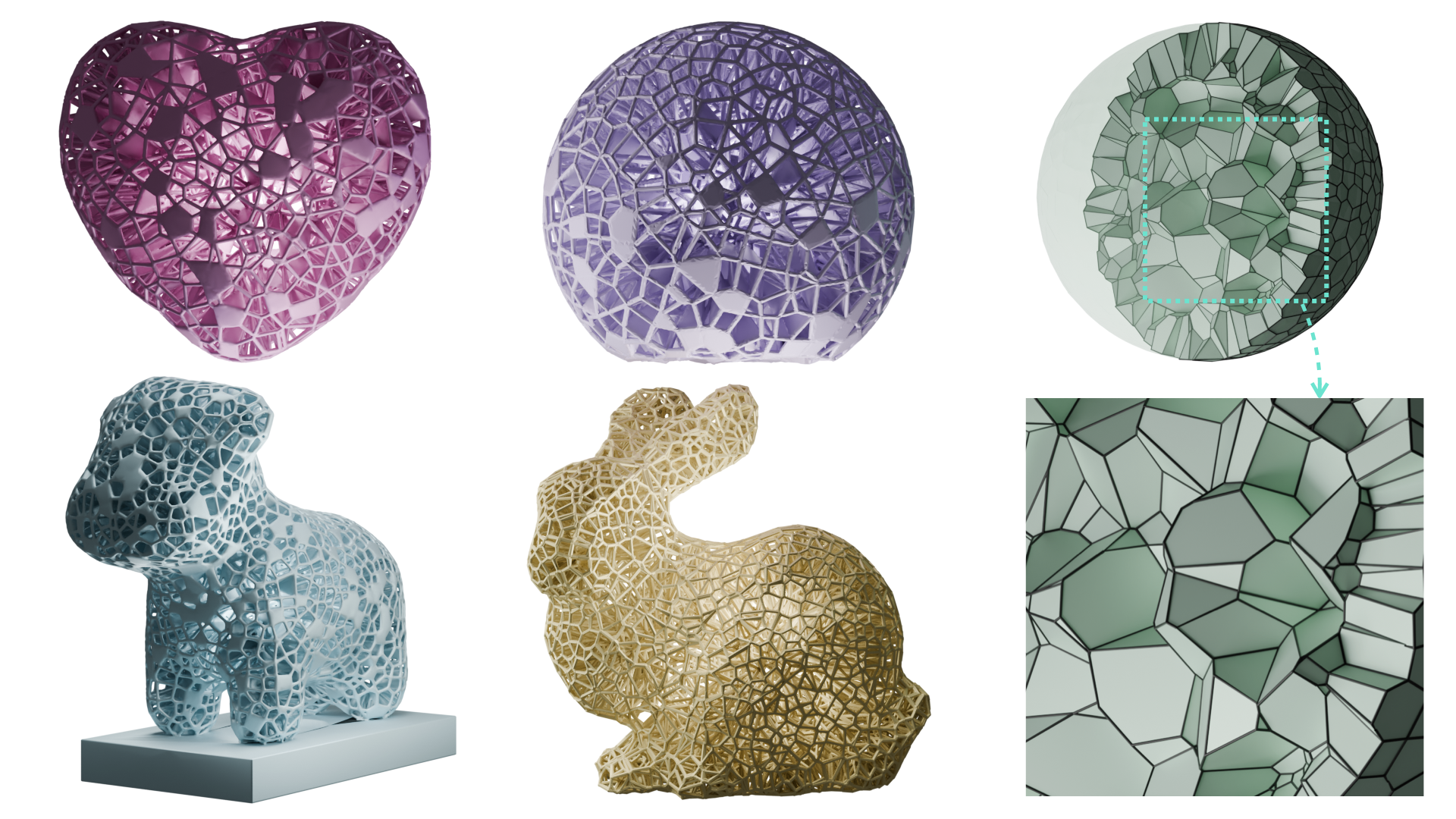}
\end{center}

\refstepcounter{figure}
\begin{flushleft}
\vspace{-10pt}
{\small \textbf{Figure \thefigure:}
VoroLight generates smooth, watertight Voronoi surface meshes via differentiable sphere-intersection training, and naturally extends to volumetric tessellations with globally consistent topology. We demonstrate the volumetric extension through 3D-printable Voronoi lamp designs.}
\end{flushleft}
\label{fig:teaser}

\blfootnote{* equal contribution. $^{1}$ University of California, Los Angeles, $^{2}$ University of Southern California,
$^{3}$ University of Utah.
Correspondence: \textit{jiayin\_lu@math.ucla.edu, yingjiang@ucla.edu, heyumeng@usc.edu, cffjiang@ucla.edu, yin.yang@utah.edu}}

\begin{abstract}
Voronoi diagrams naturally produce convex, watertight, and topologically consistent cells, making them an appealing representation for 3D shape reconstruction. However, standard differentiable Voronoi approaches typically optimize generator positions in stable configurations, which can lead to locally uneven surface geometry.
We present \textit{VoroLight}, a differentiable framework that promotes controlled Voronoi degeneracy for smooth surface reconstruction. Instead of optimizing generator positions alone, VoroLight associates each Voronoi surface vertex with a trainable sphere and introduces a sphere-intersection loss that encourages higher-order equidistance among face-incident generators. This formulation improves surface regularity while preserving intrinsic Voronoi properties such as watertightness and convexity.
Because losses are defined directly on surface vertices, VoroLight supports multimodal shape supervision from implicit fields, point clouds, meshes, and multi-view images. By introducing additional interior generators optimized under a centroidal Voronoi tessellation objective, the framework naturally extends to volumetric Voronoi meshes with consistent surface–interior topology.
Across diverse input modalities, VoroLight achieves competitive reconstruction fidelity while producing smoother and more geometrically regular Voronoi surfaces. Project page: \href{https://jiayinlu19960224.github.io/vorolight/}{https://jiayinlu19960224.github.io/vorolight/}
\end{abstract}    
\section{Introduction}
\label{sec:intro}

Reconstructing 3D shape surfaces from diverse data sources remains an important problem in computer vision and geometry processing.
Recent advances have enabled methods that generate surface meshes from multi-view images~\cite{hunyuan3d22025tencent, guo2024tetsphere, lin2023magic3d}, point clouds~\cite{maruani2023voromesh, ge2023point2mm}, and text prompts~\cite{long2024wonder3d, tsalicoglou2023textmesh}. 
Existing data-driven mesh reconstruction methods can largely be grouped into two paradigms. One line of work models geometry using neural implicit functions and recovers surfaces by extracting their zero level sets~\cite{meshSDF2020, meshUDF2022, hunyuan3d22025tencent}. The other predicts mesh connectivity and vertex positions directly, producing explicit surface representations~\cite{NDC2022, NMC2021, shen2021dmtet}. Implicit formulations excel at representing complex geometry but require an additional surface extraction step at test time, which can become computationally demanding at high resolutions. Furthermore, because the surface is defined only implicitly, enforcing geometric properties directly on the resulting mesh is nontrivial. On the other hand, explicit mesh prediction provides direct access to surface elements, yet many such methods depend on predefined regular grids or template-based discretizations, which can restrict geometric flexibility and do not inherently guarantee watertight, manifold, and self-intersection–free meshes.
Learning-based volumetric approaches, such as deformable tetrahedral representations~\cite{guo2024tetsphere} and hybrid implicit–explicit extraction methods~\cite{shen2021dmtet,chen2023flexicubes}, aim to provide interior connectivity, but typically rely on predefined grids or local primitives whose connectivity is template-dependent and does not naturally yield a globally consistent convex partition aligned with the reconstructed surface.

To address these limitations, Voronoi diagrams offer an alternative geometric representation producing convex, watertight, and topologically consistent cells~\cite{okabe09}. Recent work has explored differentiable formulations of Voronoi structures, enabling gradient-based optimization for data-driven reconstruction. \textit{VoroMesh}~\cite{maruani2023voromesh} optimizes generator positions to construct a Voronoi surface mesh that conforms to an input point cloud. In this formulation, mesh surfaces are defined as Voronoi faces between oppositely labeled generators, corresponding exactly to the bisector planes between generator points. VoroMesh introduces a \textit{VoroLoss} that approximates point-to-bisector distances, enabling gradient-based training without explicitly constructing the full Voronoi diagram. The resulting surface is guaranteed to be watertight and free of self-intersections. However, the reconstructed surfaces can be locally uneven: standard Voronoi configurations lack the higher-order intersections required for smooth curvature, causing locally bumpy geometry. Moreover, the framework is restricted to surface point-cloud inputs, limiting its adaptability to multi-view images, implicit fields, and surface meshes.

In this work, we propose VoroLight, to our knowledge the first differentiable framework that explicitly learns higher-order Voronoi degeneracy to control surface curvature.
Inspired by the sphere-based Voronoi construction of \textit{VoroCrust}~\cite{vorocrust}, which achieves smooth boundary-conforming Voronoi cells by placing generator pairs at intersection points of constructed spheres centered at mesh vertices, we extend this geometric insight into a fully differentiable framework.
We associate a trainable sphere with each Voronoi surface vertex and optimize a sphere-intersection loss that enforces all spheres incident to a face to pass through the same two intersection points.
Because such configurations arise only under higher-order equidistance, the constraint promotes controlled degeneracy and smooth, well-regularized surface geometry (Sec.~\ref{sec:surface_sphere_refinement}).
Because losses are defined directly on surface vertices, VoroLight supports multimodal shape supervision. The formulation also extends naturally to volumetric Voronoi meshes by introducing interior generators without altering the optimized surface.
Volumetric Voronoi diagrams have proven useful in numerical simulations such as fluid dynamics~\cite{VoroFEMfluid2010,de2015power} and solid mechanics~\cite{VoroFEM2012Solic, liu2018voronoi}.

Our contributions are: (a). VoroLight, a differentiable Voronoi surface formulation that learns higher-order degeneracy via trainable vertex spheres and a sphere-intersection loss, enabling smooth and watertight surface reconstruction. (b). A unified framework that supports multimodal shape supervision. (c). A natural extension to volumetric Voronoi meshes with consistent surface–interior topology, demonstrated through reconstruction experiments and 3D-printable Voronoi lamp fabrication. (d). Comprehensive evaluations across diverse input modalities demonstrate improved surface regularity, competitive reconstruction fidelity, and robustness under sparse or noisy supervision.

\section{Related Work}
\label{sec:related}

\subsection{Surface and Volumetric Mesh Reconstruction}
\label{sec:related_surface_volumetric}

Learning-based 3D surface reconstruction methods generate explicit meshes from diverse inputs, including multi-view images~\cite{liu2023one,long2024wonder3d,zhao2025hunyuan3d}, point clouds~\cite{hanocka2020point2mesh,ge2023point2mm}, and implicit fields~\cite{wang2021neus}. While these approaches can capture fine geometric detail by directly predicting triangle meshes, they often lack geometric guarantees such as watertightness, manifoldness, and freedom from self-intersections, limiting their suitability for downstream tasks requiring physically valid meshes. As noted in the survey article~\cite{MeshGenSurvey2022}, generating watertight manifold meshes remains a significant open challenge. Volumetric reconstruction methods further aim to produce meshes with explicit interior connectivity. Learning-based approaches such as \textit{TetSphere}~\cite{guo2024tetsphere} represent shapes using deformable tetrahedral primitives; however, the resulting meshes may contain overlapping elements, surface gaps, and inconsistent interior topology, which hinder downstream applications that require physically valid meshes. Classical meshing tools such as \textit{TetGen}~\cite{Si2015TetGen} generate high-quality tetrahedralizations from clean boundary meshes with strong geometric guarantees. However, \textit{TetGen} is non-differentiable, requires clean mesh inputs, and cannot be directly applied to multimodal data such as raw point clouds or multi-view images.

\subsection{Neural Implicit and Hybrid Representations}
\label{sec:related_implicit_hybrid}

Neural implicit representations model shapes as continuous fields, such as DeepSDF~\cite{park2019deepsdf}, Occupancy Networks~\cite{mescheder2019occupancy}, and NeuS~\cite{wang2021neus}, providing compact and expressive geometry priors. Volume rendering approaches, including NeRF~\cite{mildenhall2020nerf} and successors such as 3D Gaussian Splatting~\cite{kerbl3Dgaussians}, achieve high-quality novel view synthesis but produce intermediate representations (e.g., density fields or Gaussian splats) that are not directly suitable as explicit, simulation-ready meshes. Surfaces are typically extracted as zero level sets via discretization methods such as Marching Cubes~\cite{Lorensen:1987:MCA}, introducing an additional test-time step whose cost grows with desired resolution, and that can yield irregular triangle meshes and resolution-dependent artifacts. To reduce extraction artifacts, hybrid techniques combine implicit fields with differentiable surface extraction. For example, DMTet~\cite{shen2021dmtet} couples neural SDFs with differentiable marching tetrahedra for end-to-end optimization over tetrahedral meshes, while FlexiCubes~\cite{chen2023flexicubes} improves isosurface quality through adaptive grid-based schemes. However, these approaches generally rely on background grids or fixed templates whose connectivity does not naturally adapt to complex geometry, limiting geometric flexibility and direct control over discrete mesh elements.

\subsection{Voronoi and Differentiable Voronoi Methods}
\label{sec:related_voronoi_classical}

Voronoi diagrams partition space into convex, non-overlapping polyhedral cells with orthogonal face–edge structure—properties advantageous for numerical simulation~\cite{vorocrust,xiao2018optimal,maruani2023voromesh}. When constructed between oppositely labeled generators, Voronoi faces enable watertight, non-self-intersecting surface reconstruction from unstructured inputs~\cite{maruani2023voromesh}. However, standard Voronoi diagrams are unbounded and do not inherently conform to geometric boundaries, often requiring ad-hoc clipping that compromises convexity. 
VoroCrust~\cite{vorocrust} enforces boundary conformity through a provably correct sphere-based construction that places generator pairs at triple-sphere intersections so their bisector planes reproduce input facets. While it produces high-quality boundary-aligned Voronoi cells, it is deterministic, non-differentiable, and assumes clean triangle meshes as input.

Recent work integrates Voronoi geometry into differentiable frameworks for data-driven reconstruction. \textit{VoroMesh}~\cite{maruani2023voromesh} optimizes generator positions using a \textit{VoroLoss} that approximates point-to-bisector distances, enabling gradient-based training to produce watertight meshes from point clouds without explicitly constructing the full Voronoi diagram. However, it relies on point-cloud supervision and lacks explicit surface regularization, often leading to uneven surface geometry. Other differentiable formulations derive closed-form expressions for Voronoi vertices with respect to generator positions~\cite{numerow2024differentiable}, enabling Newton-type optimization for cellular and foam simulations. These implicit formulations support continuous topology changes (e.g., cell splitting and merging) without re-meshing or contact handling, but they are not designed to regularize surface smoothness under multi-modal supervision. 
In contrast, VoroLight introduces a differentiable sphere-intersection parameterization that enforces controlled Voronoi degeneracy at surface vertices, extending VoroCrust’s geometric insight into a trainable framework. This design regularizes surface geometry and enables direct multimodal supervision on mesh vertices, addressing key limitations of prior differentiable Voronoi methods. By adding interior generators, the framework naturally extends to volumetric Voronoi meshes with consistent surface–interior topology.

\section{Method}
\label{sec:method}

We first analyze stable and degenerate Voronoi configurations, then describe initialization and sphere-intersection training, followed by volumetric extension.

\subsection{Stable and Degenerate Voronoi Configurations}
\label{sec:initial_surface}

\begin{figure*}[t]
	\centering
	\includegraphics[width=0.5\linewidth]{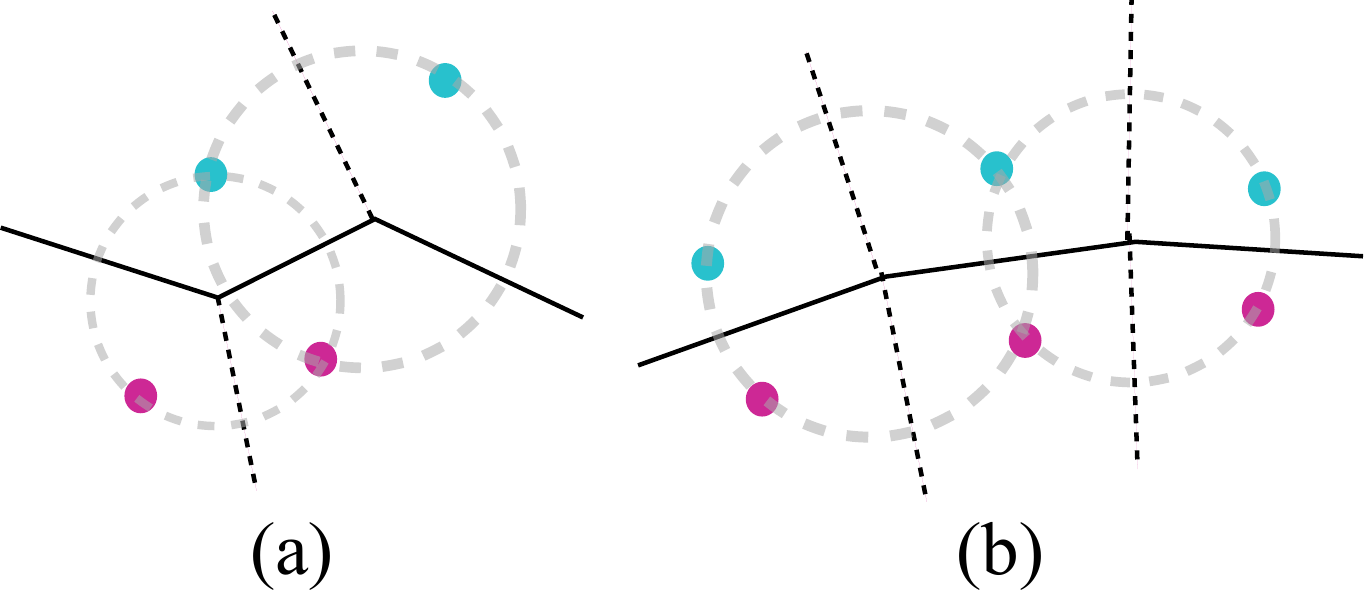}
	\caption{Schematic in 2D of (a) a \emph{stable} Voronoi vertex, defined by the intersection of three generator bisectors, and (b) a \emph{degenerate} Voronoi vertex, defined by four or more equidistant generators. Degenerate configurations are necessary for smooth, continuously curved Voronoi surfaces.}
	\label{fig:stable_degenerate_voro_2d}
		\vspace{-5pt}
\end{figure*}

A Voronoi vertex is a point equidistant from a set of generators. In 3D, a \emph{stable} vertex is defined by exactly four generators and remains under small perturbations of the generators. A \emph{degenerate} vertex is shared by five or more generators; this non-generic configuration typically splits into multiple stable vertices under generator perturbations.
As illustrated by the 2D analogy in Fig.~\ref{fig:stable_degenerate_voro_2d}, stable vertices produce jagged boundaries, whereas smooth arcs require degenerate configurations. 
The same holds in 3D. 
Surfaces composed of stable vertices exhibit locally uneven geometry, whereas higher-order degenerate vertices couple neighboring face orientations and enable smooth curvature. 
Existing differentiable Voronoi methods optimize stable configurations and do not enforce such degeneracy.
VoroLight explicitly trains for it via the sphere-based parameterization described in Sec.~\ref{sec:surface_sphere_refinement}.

\begin{figure*}[t]
	\centering
	\includegraphics[width=\linewidth]{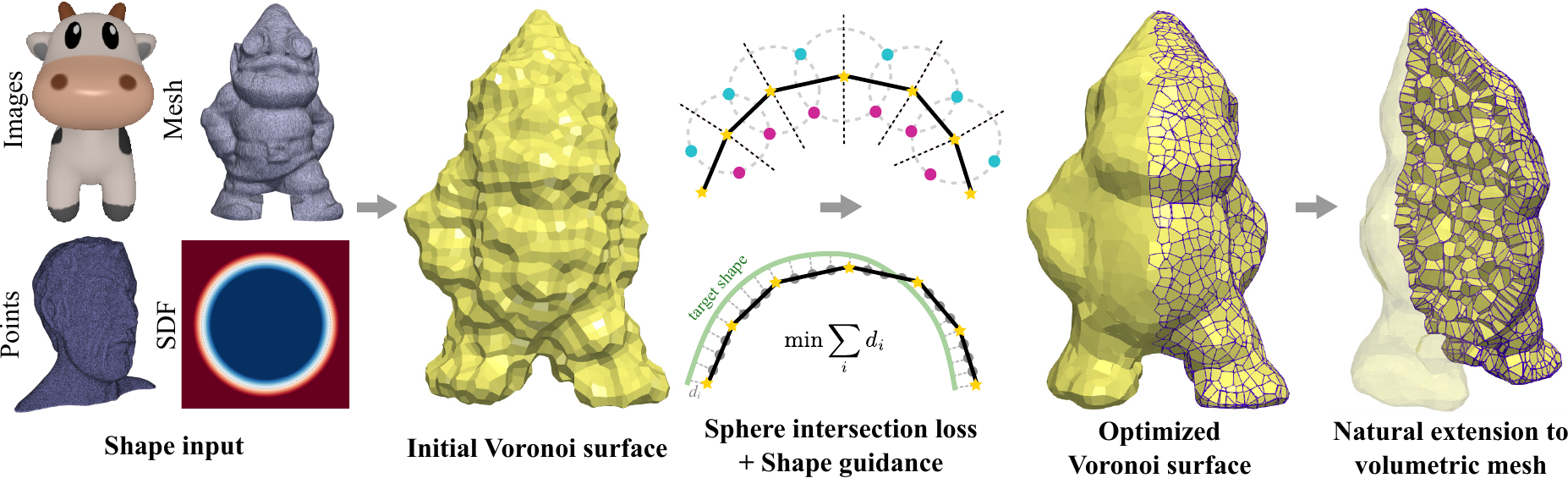}
	\caption{
		\textbf{\textit{VoroLight} pipeline.}
		VoroLight pipeline. An initial rough Voronoi surface is constructed from diverse inputs and refined via \emph{sphere-intersection} training, which enforces controlled degeneracy and smooths Voronoi surface. Interior generators extend the surface to a volumetric Voronoi mesh. Example: $n_{\text{init}}=25$, $N_{\text{inner}}=800$.
	}
	\label{fig:pipeline}
	\vspace{-5pt}
\end{figure*}

\subsection{Mesh Initialization}
\label{sec:mesh_init}

We first produce an approximate Voronoi surface of the target shape using a boundary-reflection strategy.
We construct a uniform voxel grid at resolution $n_{\text{init}}$ and classify each voxel as interior or exterior of the shape.
This classification is modality-agnostic: for implicit inputs we evaluate the level-set field; for image inputs we project samples through silhouettes; for point cloud or mesh inputs we use ray casting or proximity tests.
Exterior generators $\mathcal{G}_{\text{ext}}$ are placed at voxel centers in a thin band outside the shape, and corresponding interior generators $\mathcal{G}_{\text{in}}$ are obtained by reflecting each exterior generator inward across the estimated surface normal:
\begin{equation}
    \bm{g}_{\text{in}} = \bm{g}_{\text{ext}} - 2\delta \cdot \bm{n}(\bm{g}_{\text{ext}}),
    \label{eq:reflection}
\end{equation}
where $\delta$ is the distance to the boundary and $\bm{n}$ is a Gaussian-smoothed surface normal. 
The Voronoi surface is extracted between oppositely labeled generators. Using a differentiable Voronoi formulation~\cite{numerow2024differentiable} that expresses each Voronoi vertex as a closed-form, differentiable function of the generator positions, we further improve the surface by minimizing
\[
\mathcal{L}_{\text{init}} = \mathcal{L}_{\text{shape}} + \mathcal{L}_{\text{reg}},
\]
where $\mathcal{L}_{\text{shape}}$ is a shape conforming loss and $\mathcal{L}_{\text{reg}}$ is a surface regularization loss, as described in Sec.~\ref{sec:losses}.

This process yields an initial Voronoi surface composed of uniformly sized polygonal faces that approximate the target geometry and provide a suitable initialization for sphere intersection training. Full implementation details in Appendix. However, as discussed in Sec.~\ref{sec:initial_surface}, the mesh remains confined to stable configurations and is locally bumpy, a limitation addressed by sphere-intersection training.

\subsection{Sphere Intersection Training Scheme}
\label{sec:surface_sphere_refinement}

We introduce a differentiable sphere-intersection formulation that promotes degenerate Voronoi configurations at surface vertices.

\paragraph{From VoroCrust to a differentiable formulation.}
VoroCrust~\cite{vorocrust} constructs smooth, boundary-conforming Voronoi meshes by placing generator pairs at the intersection loci of sphere triples centered at surface samples. 
The bisector planes of such generator pairs enforce degenerate higher-order equidistance and reproduce the input surface facets.
However, VoroCrust is a deterministic, non-differentiable algorithm that requires a clean triangle mesh as input.

VoroLight lifts this geometric insight into a fully differentiable, gradient-based framework. Instead of deterministically computing sphere placements from a fixed mesh, we associate each Voronoi surface vertex with a \emph{trainable sphere} and jointly optimize sphere parameters and generator positions under
\[
\mathcal{L}_{\text{train}} = \mathcal{L}_{\text{sphere}} + \mathcal{L}_{\text{shape}} + \mathcal{L}_{\text{reg}},
\]
where $\mathcal{L}_{\text{sphere}}$ is the sphere-intersection loss that enforces degeneracy, as required for a locally smooth Voronoi surface (Sec.~\ref{sec:initial_surface}). $\mathcal{L}_{\text{shape}}$ is a target shape conforming loss that adapts to the type of input modality. 
This formulation enables learning a smooth Voronoi surface that conforms to the target geometry across diverse input modalities.

\begin{wrapfigure}{t}{0.25\textwidth}
	\vspace{-1em}
	\centering
	\includegraphics[width=\linewidth]{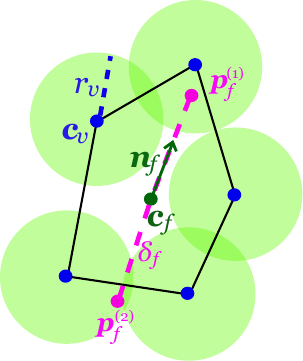}
	\caption{Face sphere-intersection parameterization.
		Each face $f$ defines symmetric targets $\bm p_f^{(1)}, \bm p_f^{(2)}$ via $\bm c_f$ and $\bm n_f$; all incident spheres $(\bm c_v, r_v)$ are
		constrained to pass through both points.}
	\label{fig:sphere_params_schematic}
	\vspace{-0.5em}
\end{wrapfigure}

\paragraph{Sphere parameterization.}
As shown in Fig.~\ref{fig:sphere_params_schematic},
for each surface vertex $v$, we define a trainable sphere $S_v = (\bm{c}_v, r_v)$ with center $\bm{c}_v$ initialized at the initial surface mesh vertex position and radius $r_v$ initialized as the average distance from the vertex to the centroids of its incident faces. 
For each surface face $f$, we introduce a scalar learnable offset $\delta_f$, initialized as $\sqrt{A_f}$, where $A_f$ is the face area. It defines two \emph{target intersection points}:
\begin{equation}
    \bm{p}_f^{(1)} = \bm{c}_f + \delta_f\,\bm{n}_f, \qquad
    \bm{p}_f^{(2)} = \bm{c}_f - \delta_f\,\bm{n}_f,
    \label{eq:face_intersections}
\end{equation}
where $\bm{c}_f$ and $\bm{n}_f$ are the face centroid and unit outward normal.
Placing both intersection points symmetrically along the face normal keeps them on the perpendicular bisector plane of the face generators, discouraging skewed configurations and promoting convex, well-shaped faces.
The trainable parameters are $\{\bm{c}_v, r_v\}_{v}$ and $\{\delta_f\}_{f}$. $\mathcal{L}_{\text{sphere}}$ enforces that all spheres incident to face $f$ intersect at exactly two points, the point pair $\bm{p}_f^{(1)}$ and $\bm{p}_f^{(2)}$.

\paragraph{Polygon-face sphere constraints.}
We apply sphere-intersection constraints directly on the original polygon faces of the Voronoi surface. 
For a face with $n$ incident vertices, this requires $n$ spheres to pass through the same two target intersection points. 
For $n>3$, the resulting system is globally over-constrained, coupling all face-incident spheres through shared intersection targets. 
In practice, optimization converges to a stable near-consistent configuration with small residual deviations, which may manifest as minor face misalignments in regions of high curvature or sharp features. 
Despite this mild approximation, the formulation effectively promotes higher-order degeneracy and smooth, well-regularized surface geometry.

\paragraph{Triangle-face relaxation (ablation).}
As an optional relaxation, each polygon face can be subdivided into triangles by connecting its vertices to the face centroid. 
Each triangular face then involves exactly three incident spheres, yielding a locally consistent sphere system whose intersection constraints admit near-zero residual solutions. 
We study this relaxation in Sec.~\ref{sec: ablation} to analyze the effect of constraint consistency on surface quality.

\subsection{Sphere Intersection Loss}
\label{sec:sphere_loss}

The sphere intersection loss includes two components, $\mathcal{L}_{\text{sphere}} = \mathcal{L}_{\text{int}} + \mathcal{L}_{\text{excl}}$, expressed via the power distance $g(\bm{p}, \bm{c}, r) = \|\bm{p} - \bm{c}\|^2 - r^2$.

\paragraph{Intersection loss.}
All spheres incident to face $f$ must pass through both target intersection points:
\begin{equation}
    \mathcal{L}_{\text{int}}
    = \frac{1}{2N_{\text{pairs}}} \sum_{f}\;\sum_{v\in V(f)} \sum_{k=1,2}
    \bigl[g(\bm{p}_f^{(k)},\, \bm{c}_v, r_v)\bigr]^{2},
    \label{eq:intersection_loss}
\end{equation}
where $V(f)$ is the set of vertices incident to face $f$, and $N_{\text{pairs}} = \sum_f |V(f)|$.

\paragraph{Exclusion loss.}
Spheres \emph{not} incident to face $f$ must not enclose the target intersection points:
\begin{equation}
    \mathcal{L}_{\text{excl}}
    = \frac{1}{N_{\text{viol}}} \sum_{f}\;\sum_{v\notin V(f)} \sum_{k=1,2}
    \bigl[\max\bigl(0,\, m - g(\bm{p}_f^{(k)},\bm{c}_v,r_v)\bigr)\bigr]^{2},
    \label{eq:exclusion_loss}
\end{equation}
where $m = 10^{-13}$ and $N_{\text{viol}}$ counts active violations.
Non-incident sphere queries are accelerated with a KD-tree.

The exclusion loss is necessary because $\mathcal{L}_{\text{int}}$ alone does not prevent a non-incident sphere from containing an intersection point (i.e., $g < 0$), which would be geometrically inconsistent; $\mathcal{L}_{\text{excl}}$ explicitly penalizes such violations.

\subsection{Loss Functions}
\label{sec:losses}

\paragraph{Shape losses $\mathcal{L}_{\text{shape}}$.}
We define modality-specific losses that align the Voronoi surface with the target geometry.
For \emph{implicit inputs} (e.g., signed distance fields), we minimize the squared level-set error at surface samples, normalized by the characteristic length scale $h_0^2$.
For \emph{point cloud and mesh inputs} (mesh surfaces are converted to a dense point cloud by uniform sampling), we minimize the bidirectional Chamfer distance between Voronoi surface samples and target points.
For \emph{multi-view image inputs}, we minimize silhouette mask, surface normal map, and neural shading losses using the differentiable rasterizer nvdiffrast~\cite{Laine2020diffrast} and a neural deferred shader (NDS)~\cite{worchel2022nds}.
Full shape loss definitions are in Appendix.

\paragraph{Regularization losses $\mathcal{L}_{\text{reg}}$.}
To promote smooth, well-shaped surfaces we apply three regularizers, all normalized by the mean face area $\bar{A}$:
\emph{normal consistency} $\mathcal{L}_{\text{normal}}$ penalizes angular deviation between a face normal and the average of its neighbors, \emph{area uniformity} $\mathcal{L}_{\text{area}}$ penalizes face-area variance, and 
\emph{face centroid alignment $\mathcal{L}_{\text{fc}}$} penalizes the distance between each face centroid and the average centroid of its neighboring faces.
During mesh initialization, all three losses are active; during sphere intersection training, only $\mathcal{L}_{\text{normal}}$ is applied.
Full definitions are in Appendix.

\subsection{Volumetric Mesh Extension}
\label{sec:volumetric_mesh}

The optimized surface generators can be extended to a full volumetric Voronoi tessellation. 
We introduce $N_{\text{inner}}$ additional generators placed deep inside the shape---positioned far enough from the surface that their Voronoi cells cannot intersect the surface boundary and therefore leave the optimized surface mesh entirely unchanged.

\paragraph{Voronoi-verified interior point sampling.}
Naively placing points inside the shape does not guarantee the non-interference property: even a geometrically interior point can have a Voronoi cell that reaches the surface when combined with the existing surface generators.
We therefore employ a \emph{Voronoi-verified iterative sampling} strategy: we sample candidate points, compute the full Voronoi diagram together with all surface generators, and iteratively remove any candidate whose Voronoi cell intersects the surface boundary.
This process repeats until $N_{\text{inner}}$ valid deep interior generators are accumulated. Full details in Appendix.

\paragraph{CVT optimization.}
With all surface generators fixed, we optimize only the deep interior generators $\mathcal{G}_{\text{inner}}$ under the centroidal Voronoi tessellation (CVT) loss~\cite{du99} using the differentiable Voronoi diagram formulation~\cite{numerow2024differentiable} mentioned in Sec.~\ref{sec:mesh_init}.
We minimize:
\begin{equation}
    \mathcal{L}_{\text{CVT}}
    = \frac{1}{N_{\text{inner}}}\sum_{i\in\mathcal{G}_{\text{inner}}}
      \frac{\|\bm{p}_i - \bar{\bm{c}}_i\|^{2}}{\bar{A}},
\end{equation}
where $\bm{p}_i$ is the generator position and $\bar{\bm{c}}_i$ is the centroid of its Voronoi cell.
This encourages each generator to coincide with its cell centroid, producing a uniform volumetric tessellation in the interior. 
This yields a watertight volumetric Voronoi mesh with globally consistent surface–interior connectivity.

\section{Experiments}
\label{sec:experiments}

We demonstrate the versatility of \textit{VoroLight} across diverse input modalities: implicit shape level-set fields, point clouds and surface meshes, and multi-view images.
We construct a testing dataset comprising $22$ object meshes: $10$ meshes without texture information randomly selected from Thingi10K~\cite{zhou2016thingi10k}, and $12$ meshes with texture information, including $11$ randomly selected from Objaverse~\cite{deitke2023objaverse} and the Spot model~\cite{crane2013robust}. We use PyTorch3D~\cite{ravi2020accelerating} and neural deferred shader (NDS)~\cite{worchel2022nds} to extract dense surface target point clouds and multi-view RGB images and surface normal maps under consistent camera poses. We further develop a Python binding for the multi-threaded \textit{Voro++} library~\cite{Rycroft2009VoroPP,LU2023multithreadedVpp} to efficiently compute Voronoi diagrams within our pipeline. For each experiment, we evaluate reconstruction quality using metrics that assess both shape fidelity (SDF error, F1 score, and Chamfer distance) and surface smoothness (mean curvature). Detailed metric formulations are provided in Appendix. Unless otherwise stated, CD$^2$ and F1 are reported as mean $\pm$ standard deviation across shapes in the evaluation set. Curvature statistics based on $|H|$ are computed per shape and then averaged across shapes. Additional application examples, including single-view 3D reconstruction and artistic 3D-printed lamp designs, are shown in Appendix.

\subsection{Implicit Shape Level-Set Field}

\begin{figure}[t]
    \centering
    \includegraphics[width=1\linewidth]{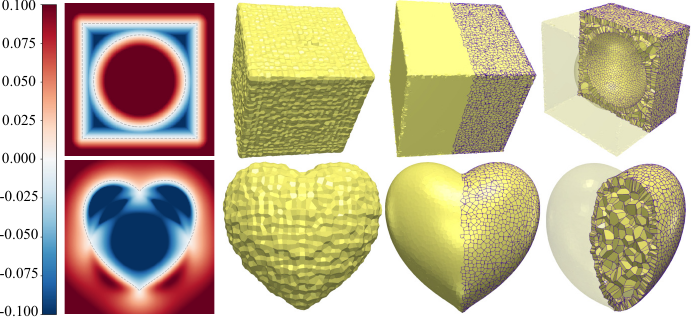}
    \caption{Reconstruction from implicit level-set fields. From left to right: input field slice; initial rough Voronoi surface; optimized smooth Voronoi surface after sphere-intersection training, partially overlaid with Voronoi edges; and volumetric Voronoi extension. Example uses $n_{\text{init}}=25$, $N_{\text{inner}}=800$.}
    
    \label{fig:taubin_heart_voronoi}
    \vspace{-5pt}
\end{figure}

We validate \textit{VoroLight} on analytic implicit fields, whose zero level set defines continuous surfaces. We test two implicit shapes: the Taubin heart~\cite{TaubinHeart1993}, and a cube with a spherical cavity constructed via signed distance field (SDF) composition.

The Taubin heart is defined as
\[
F(x,y,z) = (x^2 + \tfrac{9}{4}y^2 + z^2 - 1)^3 - x^2z^3 - \tfrac{9}{80}y^2z^3 = 0.
\]
The cube–sphere shape is formed using
\[
F(x,y,z) = \max(F_{\text{cube}}, -F_{\text{sphere}}),
\]
with 
$F_{\text{cube}}(x,y,z) = \|(|x|, |y|, |z|)\|_{\infty} - a$ and $F_{\text{sphere}}(x,y,z) = \sqrt{x^2 + y^2 + z^2} - r$.

\begin{table}[t]
    \centering
    \small
    \setlength{\tabcolsep}{4pt}
    \begin{tabular*}{1\linewidth}{@{\extracolsep{\fill}}l cccc@{}}
        \toprule
        Shape & Mean $|H|$ $\downarrow$ & Median $|H|$ $\downarrow$ & P95 $|H|$ $\downarrow$ & SDF Loss $\downarrow$ \\
        \midrule
        Cube Before  & 10.5274 $\pm$ 9.927 & 7.3243 & 25.3509 & 0.0055 \\
        Cube After   & 1.6657 $\pm$ 4.4819 & 0.0141 & 7.5055 & 0.0002 \\
        \midrule
        Heart Before & 4.4406 $\pm$ 3.2158 & 4.119 & 10.2855 & 0.0091 \\
        Heart After  & 1.4836 $\pm$ 0.8593 & 1.3428 & 2.5150 & 0.0013 \\
        \bottomrule
    \end{tabular*}
    \caption{Curvature-based smoothness statistics and SDF loss before and after optimization for the cube and heart shapes. Lower values indicate smoother surfaces for the curvature-based metrics.}
    \label{tab:curvature_smoothness}
        \vspace{-5pt}
\end{table}

As shown in Fig.~\ref{fig:taubin_heart_voronoi}, starting from a rough initial Voronoi mesh surface, \textit{VoroLight} produces smooth, watertight Voronoi surfaces and consistent volumetric Voronoi tessellations.
Along sharp edges, slight Voronoi face misalignments appear because face spheres must simultaneously satisfy constraints from adjacent faces with discontinuous normals. With uniform generator density, the optimizer converges to a compromise rather than an exact crease.
A feature-aware, adaptively refined generator distribution along sharp edges or high-curvature regions could address this limitation in future work.
For smooth regions, the sphere intersection training reliably recovers well-regularized, shape-conforming surfaces directly from the implicit field. For quantitative evaluation, we compare the reconstructed surfaces before and after sphere-intersection training on both the cube-with-cavity and the Taubin heart. We report curvature-based smoothness statistics together with an SDF consistency term, as summarized in Table~\ref{tab:curvature_smoothness}. As shown, sphere-intersection training with SDF shape supervision consistently reduces all curvature-based metrics and the SDF loss for both test shapes, confirming that the optimized Voronoi surfaces become smoother while becoming better aligned with the target implicit geometry.

\subsection{Point Cloud and Surface Mesh}
\label{sec:surface mesh and pt cld}

\begin{figure}[t]
    \centering
    \includegraphics[width=1\linewidth]{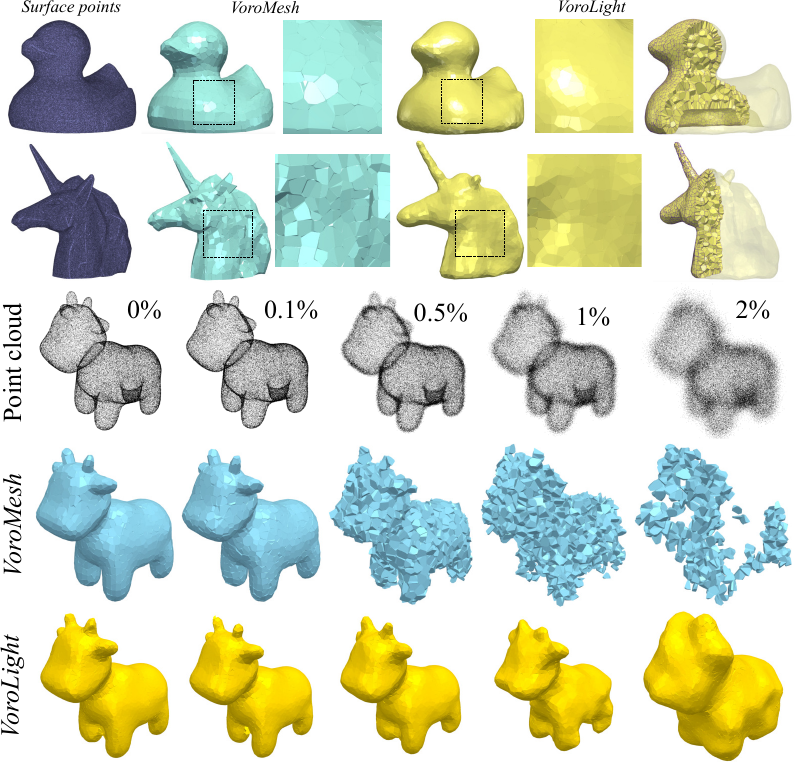}
    \caption{Comparison with VoroMesh on surface and noisy point-cloud reconstruction. 
Top: surface point-cloud fitting ($153{,}600$ points). 
Bottom: reconstruction under increasing Gaussian noise levels (0\%–2\%). 
Example uses $n_{\text{init}}=25$, $N_{\text{inner}}=800$.}
    \label{fig:mesh_pcl_full}
\end{figure}

\begin{table}[t]
  \centering
  \begin{tabular*}{\linewidth}{@{\extracolsep{\fill}}l ccccc@{}}
    \toprule
    Method & CD$^2$ ($\times 10^{-3}$) $\downarrow$ &F1@0.01 $\uparrow$ & Mean $|H|$ $\downarrow$ & Median $|H|$ $\downarrow$ & P95 $|H|$ $\downarrow$\\
    \midrule
    \textit{VoroLight} & $0.053 \pm 0.027$ & $0.9575 \pm 0.0293$ & $ 14.3868 \pm 14.5309 $ & 10.4958 & $47.1300$\\
    \textit{VoroMesh}  & $0.015\pm 0.005$ & $0.9996 \pm 0.0007 $ & $ 26.9990 \pm 21.4034$ & 19.0778 & $74.0999$\\
    \bottomrule
  \end{tabular*}
  \caption{Comparison on 10 Thingi10K samples. VoroLight improves surface regularity with reduced curvature magnitude and fewer high-curvature outliers, while VoroMesh preserves finer geometric detail.}
  \label{tbl:comp_voromesh}
    \vspace{-5pt}
\end{table}

For input surface meshes, we reformulate reconstruction as fitting to a target point cloud. We evaluate on $10$ Thingi10K objects from our dataset, uniformly sampling dense points on each ground-truth surface to form the supervision signal. 
Reconstruction is supervised using the bidirectional Chamfer distance between points sampled on the \textit{VoroLight} surface and the target point cloud. Nearest neighbors are computed efficiently using KNN search. This point-based formulation avoids repeated potentially expensive point-to-mesh distance queries while accurately capturing surface geometry. We compare against \textit{VoroMesh}~\cite{maruani2023voromesh}. 

As shown in Fig.~\ref{fig:mesh_pcl_full}, both methods produce watertight meshes. 
\textit{VoroMesh} better preserves high-frequency geometric details, whereas \textit{VoroLight} produces smoother surfaces with more consistently aligned Voronoi faces and fewer irregular patches. 
Under increasing input noise, \textit{VoroLight} remains stable and preserves coherent cell structure, while \textit{VoroMesh} progressively fragments into irregular, disconnected cells. 
This robustness stems from the sphere-intersection coupling, which regularizes generator configurations and suppresses noise-induced local oscillations.
Moreover, \textit{VoroLight} naturally extends to full volumetric Voronoi tessellations, a capability not supported by \textit{VoroMesh}.

Quantitative results from clean inputs are reported in Table~\ref{tbl:comp_voromesh}.
\textit{VoroMesh} achieves a lower mean Chamfer distance and a higher mean F1 score, reflecting stronger recovery of fine-scale geometric detail. 
In contrast, \textit{VoroLight} reduces both the mean and 95th percentile of absolute mean curvature $|H|$, indicating fewer high-curvature outliers and improved surface regularity. 
The Chamfer gap reflects a structural tradeoff introduced by enforcing higher-order Voronoi degeneracy. 
Coupling face-incident generators through shared sphere intersections suppresses small, noise-sensitive face fragments and promotes coherent cell alignment. 
At fixed generator resolution, this yields smoother surfaces with fewer curvature outliers, while recovering less high-frequency detail. 
Increasing generator density reduces this fidelity gap (see Appendix). 
We further analyze a triangle-face relaxation with the same $n_{\text{init}}$ in Sec.~\ref{sec: ablation}, which improves reconstruction fidelity while preserving similar surface regularity. 
Additional qualitative comparisons are provided in Appendix.

\subsection{Multi-view Images}
\label{sec:multi-view img}

\begin{table}[t]
  \centering
  \small
  \setlength{\tabcolsep}{4pt} % tighten horizontal padding
  % \begin{tabular}{l cc cc}
  \begin{tabular*}{\linewidth}{@{\extracolsep{\fill}}l cc cc@{}}
    \toprule
    & \multicolumn{2}{c}{300 Views} & \multicolumn{2}{c}{6 Views} \\
    \cmidrule(lr){2-3} \cmidrule(lr){4-5}
    Method 
    & CD$^2$ ($\times 10^{-3}$) $\downarrow$ 
    & F1@0.01 $\uparrow$
    & CD$^2$ ($\times 10^{-3}$) $\downarrow$ 
    & F1@0.01 $\uparrow$ \\
    \midrule

    \textit{VoroLight} 
    & \makecell{0.289 $\pm$ 0.276 \\ (0.180)} 
    & \makecell{0.783 $\pm$ 0.139 \\ (0.809)}
    & \makecell{1.981 $\pm$ 2.312 \\ (1.199)}
    & \makecell{0.487 $\pm$ 0.168 \\ (0.555)} \\

    \textit{TetSphere} 
    & \makecell{0.953 $\pm$ 0.632 \\ (0.774)} 
    & \makecell{0.524 $\pm$ 0.127 \\ (0.520)}
    & \makecell{2.049 $\pm$ 1.770 \\ (1.748)}
    & \makecell{0.360 $\pm$ 0.121 \\ (0.338)} \\

    \bottomrule
  \end{tabular*}
  \caption{Comparison of \textit{VoroLight} and \textit{TetSphere} on multi-view image reconstruction. We report mean $\pm$ standard deviation, with median in parentheses, over 12 objects.}
  \label{tbl:comp_tetsphere}
      \vspace{-5pt}
\end{table}

\begin{figure}[t]
    \centering
    \includegraphics[width=1\linewidth]{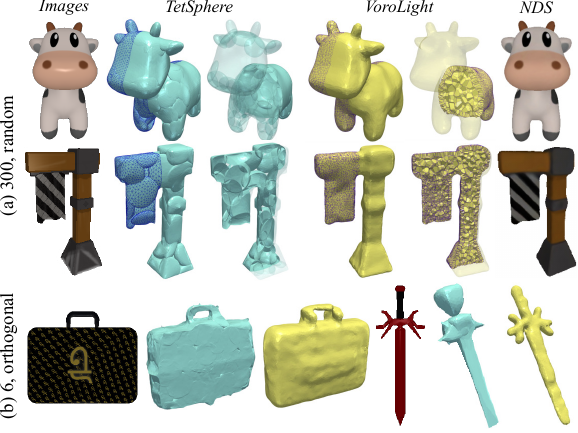}
    \caption{Comparison with \textit{TetSphere Splatting} on multi-view image reconstruction.
    Each object is reconstructed from (a) 300 random views and (b) 6 orthogonal views. 
    Examples use $n_{\text{init}} = 40$ and $N_{\text{inner}}=800$.}
    \label{fig:comp_tetsphere}
        \vspace{-5pt}
\end{figure}

We compare \textit{VoroLight} with \textit{TetSphere Splatting}~\cite{guo2024tetsphere} on multi-view image reconstruction (Fig.~\ref{fig:comp_tetsphere}).
We use the $12$ textured meshes from our dataset and render $300$ random views as well as $6$ orthogonal views (front, back, left, right, top, bottom) for each object, including corresponding depth and normal maps.
\textit{TetSphere} is trained with silhouette and depth supervision and subsequently optimized with shading to recover texture.
\textit{VoroLight} uses silhouette and normal-map supervision via nvdiffrast~\cite{Laine2020diffrast} for geometry, followed by training a neural deferred shader (NDS)~\cite{worchel2022nds} to recover appearance.
Although the two methods rely on different geometric supervision signals (depth vs.\ normal maps), both are optimized under silhouette and shading objectives.

As shown in Fig.~\ref{fig:comp_tetsphere}, \textit{TetSphere} represents shapes as collections of tetrahedral elements without explicitly enforcing a globally consistent volumetric tessellation, which can lead to overlaps or small gaps between local clusters.
In contrast, \textit{VoroLight} produces watertight volumetric meshes with globally consistent topology.
When reducing the number of input views from $300$ to $6$, reconstruction quality decreases for both methods, with a more pronounced degradation observed for \textit{TetSphere} in some cases.
Quantitatively (Table~\ref{tbl:comp_tetsphere}), \textit{VoroLight} achieves lower mean and median squared Chamfer distance and higher mean F1 score than \textit{TetSphere} under both view settings, indicating improved geometric accuracy.
Additional qualitative comparisons are provided in Appendix.

\subsection{Ablation: Triangle-Face Sphere Relaxation}
\label{sec: ablation}

\begin{figure*}[t]
    \centering
    \includegraphics[width=1\linewidth]{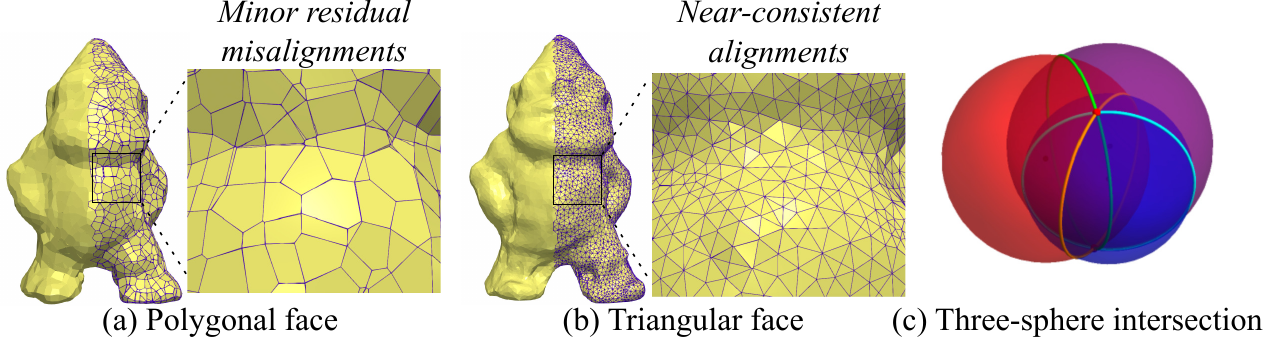}
    \caption{
    Comparison between the primary polygon-face formulation and the triangle-face relaxation.
    (a) Polygon-face constraints couple more than three spheres, forming a globally over-constrained system that may introduce minor residual misalignments.
    (b) Subdivision into triangular faces yields a locally consistent three-sphere intersection system and near-consistent alignments.
    (c) Illustration of three-sphere intersection in general position.
    }
    \label{fig:poly_tria_variant_3sphere}

    \vspace{-5pt}
\end{figure*}

The polygon-face formulation requires more than three spheres to pass through two shared intersection points, forming a globally coupled and over-constrained system. In contrast, the triangle-face relaxation subdivides each polygon into triangles by connecting face vertices to the face centroid, so that each face involves exactly three incident spheres. As illustrated in Fig.~\ref{fig:poly_tria_variant_3sphere}(c), three spheres in general position intersect along a radical line, producing exactly two consistent intersection points (one highlighted in red). The resulting constraint system is locally consistent and admits near-zero residual solutions.
As shown in Fig.~\ref{fig:poly_tria_variant_3sphere}(a)(b), this yields near-consistent face alignments while preserving overall surface geometry. Additional qualitative comparisons are provided in Appendix.

We analyze the triangle-face relaxation under the same training settings as Sec.~\ref{sec:surface mesh and pt cld}. 
The polygon-face statistics reported here correspond to the main method in Table~\ref{tbl:comp_voromesh}. 
Overall, the triangle-face relaxation reduces the fidelity gap with \textit{VoroMesh} while maintaining comparable surface regularity. 
Quantitatively, the two formulations exhibit similar area-weighted mean absolute curvature. 
The triangle-face variant shows a slightly higher mean $|H|$, but lower median, 95th percentile, and standard deviation of $|H|$, indicating fewer high-curvature outliers. 
At the same time, it improves reconstruction fidelity, achieving lower bidirectional Chamfer distance and higher F1 score. 
Additional ablations on initialization resolution and volumetric density are provided in Appendix.

\begin{table}[t]
    \centering
    \begin{tabular*}{\linewidth}{@{\extracolsep{\fill}}l ccccc@{}}
        \toprule
        Variant & CD$^2$ ($\times 10^{-3}$) $\downarrow$ & F1@0.01 $\uparrow$ & Mean $|H|$ $\downarrow$ & Median $|H|$ $\downarrow$ & P95 $|H|$ $\downarrow$  \\
        \midrule
        Polygon   &   $0.053 \pm 0.027$ & $0.9575 \pm 0.0293$ & $ 14.3868 \pm 14.5309 $ & 19.0778 & $47.1300$ \\
        Triangle &  $0.048 \pm 0.029$ &  $0.96 \pm 0.035$ &  $10.9633 \pm 8.6093$ & 8.3130 & 31.9124 \\
        \bottomrule
    \end{tabular*}
    \caption{Ablation study comparing the primary polygon-face formulation and the triangle-face relaxation over 10 Thingi10K samples. The triangle-face relaxation improves reconstruction fidelity and reduces curvature outliers, while maintaining comparable overall smoothness.}
    \label{tab:ablation_poly_tria}

    \vspace{-5pt}
\end{table}

\section{Future Work}

\textbf{Adaptive Cell Density.}
Incorporating adaptive surface meshing strategies would allow higher Voronoi cell density near sharp edges or high-curvature regions and fewer cells in flatter areas, improving geometric fidelity while maintaining curvature regularization. 
Likewise, an adaptive volumetric density field could increase interior cell resolution in regions of complex geometry, enabling smoother gradation from boundary to interior. 
These behaviors could be realized via gradient-based refinement, hierarchical subdivision, or learned density fields that condition cell placement on geometric, volumetric, or perceptual cues.

{
    \small
    \bibliographystyle{ieeenat_fullname}
    \bibliography{main}
}

\newpage
\appendix

\setcounter{page}{1}

\begin{center}
  {\LARGE\bfseries\textit{VoroLight}: Learning Voronoi Surface Meshes via Sphere Intersection\par}

  {\LARGE\bfseries Supplementary Material\par}
  		\vspace{5pt}

  \includegraphics[width=0.85\textwidth]{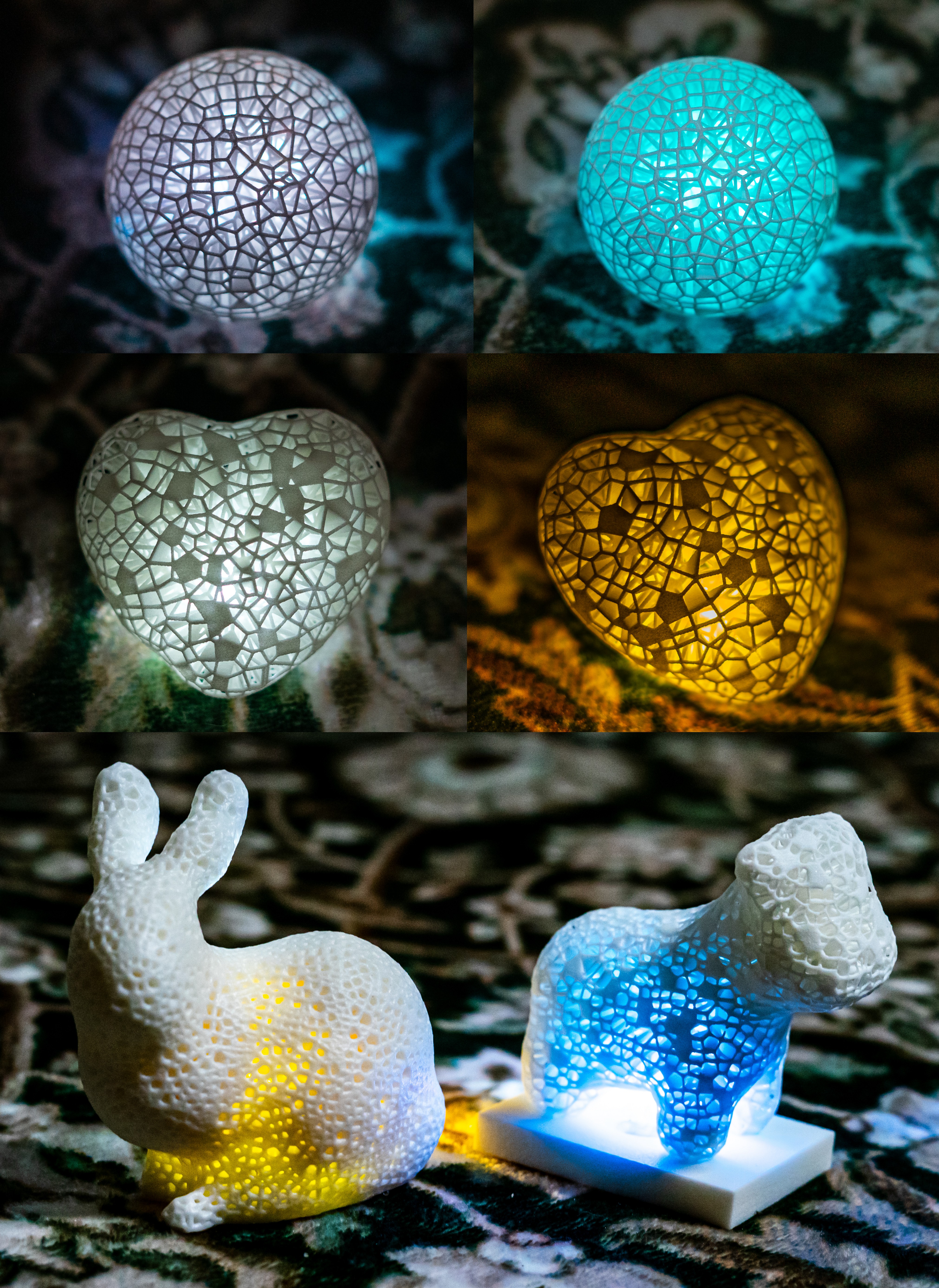}

  {\small\captionof{figure}{We designed artistic 3D printed lamps using \textit{VoroLight}: (first row) sphere lamp, (second row) heart lamp, (bottom left) bunny lamp, and (bottom right) cow lamp.}
  \label{fig:teaser_supp}}
\end{center}

\section{Detailed Loss Formulations}
\label{sec:suppl_losses}

This section provides complete mathematical formulations of all loss terms referenced in the main paper.

\subsection{Shape Losses}
\label{sec:suppl_losses_shape}

\paragraph{Implicit shape level-set field input.} Squared $L^2$ loss normalized by grid spacing~$h_0$, directly minimizing deviation from the target implicit surface:
\[
\mathcal{L}_{d} = \frac{1}{N_{\text{voro}}}\sum_{i=1}^{N_{\text{voro}}} \frac{d(\bm{x}_i)^2}{h_0^2},
\]
where $d(\bm{x}_i)$ is the level-set value at sample point~$\bm{x}_i$ on the Voronoi surface. Sample points include face vertices, face centroids, and sub-triangle centers, where sub-triangles are formed by subdividing each Voronoi face by connecting its vertices to its centroid.

\paragraph{Point cloud \& mesh surface input.}
Bidirectional Chamfer distance between Voronoi surface samples $\mathcal{P}_{\text{voro}} = \{\bm{x}_i\}$ and target samples $\mathcal{P}_{\text{target}} = \{\bm{y}_j\}$, ensuring both coverage of the target and accuracy of the reconstruction:
\begin{align}
\mathcal{L}_{\text{pc}} &=
\frac{1}{N_{\text{voro}}}\sum_{i=1}^{N_{\text{voro}}}
\min_{\bm{y}_j \in \mathcal{P}_{\text{target}}}
\|\bm{x}_i - \bm{y}_j\|^2 \nonumber\\
&\quad+
\frac{1}{N_{\text{target}}}\sum_{j=1}^{N_{\text{target}}}
\min_{\bm{x}_i \in \mathcal{P}_{\text{voro}}}
\|\bm{y}_j - \bm{x}_i\|^2.
\end{align}

\paragraph{Image input.} Combination of (1) silhouette mask loss $\mathcal{L}_{\text{mask}}$, (2) surface normal map loss $\mathcal{L}_{\text{normal}}$ via differentiable rasterizer \textit{nvdiffrast}~\cite{Laine2020diffrast}, and (3) shading loss $\mathcal{L}_{\text{shading}}$ via learned \textit{neural deferred shader (NDS)}~\cite{worchel2022nds}.

\textbf{(1) Silhouette mask loss:} Enforces correct object boundaries and silhouette shape.
\[
\mathcal{L}_{\text{mask}} = \frac{1}{N_{\text{views}}}\sum_{v=1}^{N_{\text{views}}}
\text{MSE}(M^{(v)}_{\text{target}}, M^{(v)}_{\text{render}}),
\]
where $M^{(v)}_{\text{target}}$ and $M^{(v)}_{\text{render}}$ are target and rendered binary masks, $N_{\text{views}} = \min(N_{\text{total}}, 66)$ views per iteration.

\textbf{(2) Normal map loss:}
\begin{align}
\mathcal{L}_{\text{normal}} &= 
\frac{1}{N_{\text{views}}}\sum_{v=1}^{N_{\text{views}}}
\Bigg[\sum_{(i,j)\in\Omega^{(v)}}\!
\big(1 - \langle \bm{n}^{(v)}_{\text{tgt}}(i,j),
\bm{n}^{(v)}_{\text{rnd}}(i,j)\rangle\big) \nonumber\\
&\qquad\qquad\qquad\qquad\qquad\qquad\quad
\cdot\, w^{(v)}(i,j)\Bigg].
\end{align}

where $\Omega^{(v)}$ is the set of front-facing pixels; $\bm{n}^{(v)}_{\text{tgt}}$, $\bm{n}^{(v)}_{\text{rnd}}$ are target/rendered normals; $\bm{d}^{(v)}$ is flipped view direction; weight $w^{(v)}(i,j) = \exp(|\tilde{c}^{(v)}_{\text{tgt}}(i,j)|) / Z^{(v)}$ with $\tilde{c}^{(v)}_{\text{tgt}} = c^{(v)}_{\text{tgt}}$ if $c^{(v)}_{\text{tgt}} \leq -0.1$, else $0$, and $c^{(v)}_{\text{tgt}} = \langle\bm{n}^{(v)}_{\text{tgt}},\,\bm{d}^{(v)}\rangle$; normalization $Z^{(v)} = \sum_{i,j} \exp(|\tilde{c}^{(v)}_{\text{tgt}}(i,j)|)$.

\textbf{(3) Shading loss:} Captures appearance and photometric consistency across views.
\begin{align}
\mathcal{L}_{\text{shading}} &=
\frac{1}{N_{\text{valid}}}\sum_{v \in \mathcal{V}_{\text{valid}}}
\text{SmoothL1}\big(\mathcal{F}_{\text{shader}}(\bm{p}^{(v)}, \bm{n}^{(v)}, 
\bm{d}^{(v)}_{\text{cam}}), \nonumber\\
&\qquad\qquad\qquad\quad
\bm{c}^{(v)}_{\text{target}}\big).
\end{align}

where $\mathcal{V}_{\text{valid}}$ are views with non-empty masks ($N_{\text{valid}} = |\mathcal{V}_{\text{valid}}|$); $\mathcal{F}_{\text{shader}}$ is learned \textit{NDS}~\cite{worchel2022nds}; $\bm{p}^{(v)}$, $\bm{n}^{(v)}$, $\bm{d}^{(v)}_{\text{cam}}$ are rasterized point position, normal, and view direction; evaluated on 75\% randomly sampled valid pixels per view; SmoothL1 is mean over sampled pixels across RGB channels.

\subsection{Surface Regularization Losses}
\label{sec:suppl_losses_reg}

Three regularizers normalized by mean Voronoi face area $\bar{A}$. All three are used in mesh initialization; only $\mathcal{L}_{\text{normal}}$ is used during sphere intersection training. 

\paragraph{Normal consistency.} Penalizes angular deviation between face normals and their neighborhood average, promoting smooth surface geometry:
\begin{equation}
\mathcal{L}_{\text{normal}} = \frac{1}{N_f}\sum_f \big(1 - \langle \bm{n}_f,\,\bar{\bm{n}}_f\rangle\big)^2,
\label{eq:loss_normal}
\end{equation}
where $N_f$ is total number of Voronoi faces, $\bm{n}_f$ is unit normal of face~$f$, and $\bar{\bm{n}}_f$ is normalized average of adjacent face normals.

\paragraph{Area uniformity.} Minimizes variance from mean face area, encouraging regular tessellation:
\begin{equation}
\mathcal{L}_{\text{area}} = \frac{1}{N_f}\sum_f \frac{(A_f - \bar{A})^2}{\bar{A}},
\label{eq:loss_area}
\end{equation}
where $A_f$ is the area of face~$f$.

\paragraph{Face centroid alignment.} Aligns each face centroid with neighboring face centroids, improving spatial distribution:
\begin{equation}
\mathcal{L}_{\text{fc}} = \frac{1}{N_f}\sum_f \frac{\|\bm{c}_f - \bar{\bm{c}}_f\|^2}{\bar{A}},
\label{eq:loss_centroid}
\end{equation}
where $\bm{c}_f$ is the centroid of face~$f$ and $\bar{\bm{c}}_f$ is the average of adjacent face centroids.

\section{Evaluation Metrics}
\label{app:metrics}

\subsection{CD\texorpdfstring{$^2$}{2} and F1 Score}
\label{app:metrics:cd_f1}

We uniformly sample $N = 50{,}000$ points on each surface using area-weighted triangle sampling:
\[
P_{\mathrm{gt}} = \{ \mathbf{p}_i \}_{i=1}^{N},
\qquad
P_{\mathrm{pred}} = \{ \mathbf{q}_j \}_{j=1}^{N}
\]
denote sampled point sets from ground-truth and predicted meshes.

Mean \emph{squared} bidirectional Chamfer distance CD\(^2\):
\begin{equation}
\mathrm{CD}^2
=
\frac{1}{N} \sum_{i=1}^{N} \min_{\mathbf{q}\in P_{\mathrm{pred}}}
\lVert \mathbf{p}_i - \mathbf{q} \rVert_2^2
+
\frac{1}{N} \sum_{j=1}^{N} \min_{\mathbf{p}\in P_{\mathrm{gt}}}
\lVert \mathbf{q}_j - \mathbf{p} \rVert_2^2.
\label{eq:cd2}
\end{equation}
In the main table, we report $\mathrm{CD}^2 \times 10^3$ for readability.

F1 score under threshold $\tau = 0.01$. Precision is the fraction of predicted samples whose nearest ground-truth point lies within $\tau$; recall is the fraction of ground-truth samples whose nearest predicted point lies within $\tau$:

\begin{equation}
\mathrm{Precision}
=
\frac{1}{N}\sum_{j=1}^{N}
1\!\left[
\min_{\mathbf{p}\in P_{\mathrm{gt}}}
\lVert \mathbf{q}_j - \mathbf{p} \rVert_2 < \tau
\right],
\label{eq:precision}
\end{equation}

\begin{equation}
\mathrm{Recall}
=
\frac{1}{N}\sum_{i=1}^{N}
1\!\left[
\min_{\mathbf{q}\in P_{\mathrm{pred}}}
\lVert \mathbf{p}_i - \mathbf{q} \rVert_2 < \tau
\right].
\label{eq:recall}
\end{equation}
F1 score:
\begin{equation}
\mathrm{F1}
=
\frac{2 \, \mathrm{Precision} \cdot \mathrm{Recall}}
{\mathrm{Precision} + \mathrm{Recall} + \varepsilon},
\label{eq:f1}
\end{equation}
where $\varepsilon = 10^{-12}$ is used for numerical stability.

\subsection{Curvature Metrics}
\label{app:metrics:curvature}

We estimate mean curvature $H$ at mesh vertices using local Monge-patch quadric fit. For each vertex $\mathbf{v}$, neighborhood radius
\begin{equation}
r = \rho \, h_{\mathrm{ref}}, 
\qquad \rho = 2.0,
\label{eq:curv_radius}
\end{equation}
where $h_{\mathrm{ref}}$ is median edge length of the ground-truth mesh. This radius is fixed per shape for all compared methods.

For each vertex, we collect neighboring vertices within distance $r$. Vertices with fewer than $8$ neighbors are excluded to avoid unstable local fits. We estimate vertex normal by area-weighted average of incident face normals, construct local tangent frame $(\mathbf{t}_1,\mathbf{t}_2)$, and express neighboring points in local coordinates $(x,y,z)$. We then fit quadratic height field
\begin{equation}
z = ax^2 + bxy + cy^2 + dx + ey + f
\label{eq:quadric}
\end{equation}
by weighted least squares with Gaussian weights
\begin{equation}
w(x,y,z)=\exp\!\left(-\frac{x^2+y^2+z^2}{2r^2}\right).
\label{eq:gaussian_weight}
\end{equation}
Vertices are discarded if the fitting system is rank-deficient or numerically ill-conditioned.

Given the fitted coefficients, we compute
\[
f_x=d,\qquad f_y=e,\qquad f_{xx}=2a,\qquad f_{xy}=b,\qquad f_{yy}=2c,
\]
and evaluate mean curvature using standard Monge-patch formula
\begin{equation}
H=
\frac{(1+f_y^2)f_{xx}-2f_xf_yf_{xy}+(1+f_x^2)f_{yy}}
{2(1+f_x^2+f_y^2)^{3/2}}.
\label{eq:mean_curvature}
\end{equation}
All reported curvature statistics are computed from the absolute values $|H|$ over the set of valid vertices.

Let $A_i$ denote barycentric area associated with vertex $i$ (one third of each incident triangle area), and let $\mathcal{V}_{\mathrm{valid}}$ be the set of vertices with valid curvature estimates. Normalized area weights:
\[
\tilde{A}_i = \frac{A_i}{\sum_{k\in\mathcal{V}_{\mathrm{valid}}} A_k}.
\]
From per-vertex values $\{|H_i|\}_{i\in\mathcal{V}_{\mathrm{valid}}}$, we compute surface regularity statistics including area-weighted mean absolute curvature
\begin{equation}
\overline{|H|} = \sum_{i\in\mathcal{V}_{\mathrm{valid}}} \tilde{A}_i |H_i|,
\label{eq:mean_abs_h}
\end{equation}
median and 95th percentile of $|H|$,
\begin{equation}
\mathrm{Median}(|H|),\qquad
\mathrm{P95}(|H|)=\mathrm{Percentile}_{95}\bigl(\{|H_i|\}_{i\in\mathcal{V}_{\mathrm{valid}}}\bigr),
\label{eq:p95_abs_h}
\end{equation}
and area-weighted variance and standard deviation
\begin{equation}
\mathrm{Var}(|H|)=
\sum_{i\in\mathcal{V}_{\mathrm{valid}}}
\tilde{A}_i \left(|H_i|-\overline{|H|}\right)^2,
\qquad
\mathrm{Std}(|H|)=\sqrt{\mathrm{Var}(|H|)}.
\label{eq:var_std_abs_h}
\end{equation}
Lower values indicate a smoother and more spatially regular surface.

\subsection{SDF Loss}
When applicable, SDF consistency loss:
\begin{equation}
\mathcal{L}_{\mathrm{SDF}}=
\frac{1}{N}\sum_{k=1}^{N} |\phi(\mathbf{p}_k)|,
\label{eq:sdf_loss}
\end{equation}
where $\phi$ is ground-truth signed distance field and $\mathbf{p}_k$ are uniformly sampled points on the reconstructed surface.

\section{Method Details}
\label{sec:suppl_method}

\subsection{Initialization: Boundary-Reflection Point Sampling}
\label{sec:suppl_init}

Given grid resolution parameter $n_{\text{init}}$, we identify a tight bounding box around the shape and compute grid spacing $h_0 = L_{\max} / n_{\text{init}}$ where $L_{\max}$ is the longest dimension, creating a uniform voxel grid with cubic cells of edge length $h_0$. 

\paragraph{Coarse region classification.}
We classify each voxel as interior or exterior using multi-round random sampling ($N_{\text{round}} = 35$ samples per voxel) to robustly detect thin features:
\begin{itemize}
    \item \textbf{Implicit field:} Evaluate level-set $d(\bm{p}) < 0$ for each sample; voxel is interior if any sample is inside.
    \item \textbf{Image:} Project samples to all camera views; voxel is interior if any sample is inside all silhouettes.
    \item \textbf{Mesh:} Cast rays from samples; voxel is interior if any sample's ray has odd intersection count.
    \item \textbf{Point cloud:} Check samples' proximity to surface points.
\end{itemize}
The multi-round sampling is applied in two stages:
(i) coarse tight bounding box detection on a fixed 30$^3$ grid; then
(ii) final classification on the $n_{\text{init}}$-resolution grid within the tight bounding box generates interior mask $\mathcal{M}_{\text{in}}$.
We then use morphological dilation to further obtain exterior boundary region $\mathcal{M}_{\text{ext}}$ (3-layer band outside $\mathcal{M}_{\text{in}}$).

\paragraph{Boundary-reflection point generation.}
We use a geometry-aware reflection strategy to generate well-distributed generators near the shape boundary.
First, we place exterior generators $\mathcal{G}_{\text{ext}}$ at voxel centers within $\mathcal{M}_{\text{ext}}$.
To generate corresponding interior generators $\mathcal{G}_{\text{in}}$, we compute signed distance field $d(\bm{x})$ from $\mathcal{M}_{\text{in}}$ using Euclidean distance.
We apply Gaussian smoothing ($\sigma = 1.5$ voxels) to obtain smooth surface normals:
\begin{equation}
\bm{n}(\bm{x}) = \frac{\nabla (G_{\sigma} * d)(\bm{x})}{\|\nabla (G_{\sigma} * d)(\bm{x})\| + \epsilon}, \quad \epsilon = 10^{-8}.
\end{equation}
For each exterior generator $\bm{g}_{\text{ext}}$, we compute its distance to the boundary $\delta = |d(\bm{g}_{\text{ext}})|$ and reflect it inward using the smooth surface normals:
\begin{equation}
\bm{g}_{\text{in}} = \bm{g}_{\text{ext}} - 2\delta \cdot \bm{n}(\bm{g}_{\text{ext}}),
\end{equation}
placing the interior generator at the mirror position. Reflected points are filtered to retain only those strictly inside $\mathcal{M}_{\text{in}}$.

\paragraph{Point filtering.}
An initial Voronoi diagram is computed on $\mathcal{G}_{\text{in}} \cup \mathcal{G}_{\text{ext}}$, and interior generators are filtered to retain only those whose Voronoi cells intersect the shape boundary (i.e., have faces adjacent to exterior generators). Deep interior generators that do not contribute to the surface are discarded, as Stages I and II focus solely on surface mesh optimization.

\paragraph{Differentiable Voronoi formulation.}
Following~\cite{numerow2024differentiable}, a Voronoi vertex $\bm{v}_c$ at the intersection of four generator bisector planes $\bm{x}_i$, $i \in \{0,1,2,3\}$, is computed in closed form as:
\begin{equation}
\bm{v}_c = \tfrac{1}{2}\,A^{-1}\bm{b}, \quad
A =
\begin{bmatrix}
(\bm{x}_1 - \bm{x}_0)^{\!\top} \\
(\bm{x}_2 - \bm{x}_0)^{\!\top} \\
(\bm{x}_3 - \bm{x}_0)^{\!\top}
\end{bmatrix}, \quad
\bm{b} =
\begin{bmatrix}
\|\bm{x}_1\|^2 - \|\bm{x}_0\|^2 \\
\|\bm{x}_2\|^2 - \|\bm{x}_0\|^2 \\
\|\bm{x}_3\|^2 - \|\bm{x}_0\|^2
\end{bmatrix}.
\end{equation}
This enables back-propagation from vertex-based losses to generator positions $\bm{x}_i$.
The generator set consists of interior generators $\mathcal{G}_{\text{in}}$ and exterior generators $\mathcal{G}_{\text{ext}}$.
Both $\mathcal{G}_{\text{in}}$ and $\mathcal{G}_{\text{ext}}$ positions are trainable ($N_{\text{train}} = |\mathcal{G}_{\text{in}}| + |\mathcal{G}_{\text{ext}}|$ parameters).
The Voronoi surface is extracted as faces separating $\mathcal{G}_{\text{in}}$ cells from $\mathcal{G}_{\text{ext}}$ cells.

\subsection{Deep Interior Point Generation Details}
\label{sec:suppl_deep_interior_point}

For volumetric mesh generation, we want $N_{\text{inner}}$ deep interior generator points in addition to the surface generators $\mathcal{G}_1$ (inner) and $\mathcal{G}_2$ (outer) obtained from sphere intersection training.
A critical constraint is that adding interior generators must \emph{not alter the surface mesh} that was optimized.
A naive approach of randomly sampling points inside the mesh would fail: while such points may be geometrically inside the shape, their Voronoi cells can still intersect the surface tessellation between $\mathcal{G}_1$ and $\mathcal{G}_2$, thereby modifying the optimized surface result.
We employ a \emph{Voronoi-verified iterative generation} strategy that explicitly filters out any candidate whose Voronoi cell would intersect the surface.

\paragraph{Candidate generation.}
From sphere intersection training, we obtain a triangulated surface mesh $\mathcal{M} = (\bm{V}, \bm{F})$ by subdividing each Voronoi polygon face into triangles.
We compute the mesh's tight bounding box $[\bm{m}_{\min}, \bm{m}_{\max}]$, which is typically much tighter than the global domain $[a_x, b_x] \times [a_y, b_y] \times [a_z, b_z]$, improving sampling efficiency.

At each outer iteration $k$, we generate $N_{\text{cand}}^{(k)} = \max(200, 3 \cdot N_{\text{remain}}^{(k)})$ candidate points by uniform random sampling in the mesh bounding box, where $N_{\text{remain}}^{(k)}$ is the number of points still needed.
We over-generate (3$\times$) to compensate for candidates outside the mesh and candidates that violate the fixed boundary, which will be deleted.

\paragraph{Iterative Voronoi filtering.}
For each batch of candidates $\mathcal{C}^{(k)}$, we perform an inner iterative refinement to ensure no Voronoi boundary violations remain:
\begin{enumerate}
\item Initialize $\mathcal{C}_{\text{filtered}}^{(0)} = \mathcal{C}^{(k)}$.
\item At inner iteration $j$, construct temporary generator set $\mathcal{G}_{\text{temp}}^{(j)} = \mathcal{C}_{\text{filtered}}^{(j)} \cup \mathcal{G}_1 \cup \mathcal{G}_2$ and compute Voronoi diagram.
\item Designate the first $|\mathcal{C}_{\text{filtered}}^{(j)}| + |\mathcal{G}_1|$ generators as shape points and $\mathcal{G}_2$ as exterior points. The Voronoi algorithm identifies boundary generator indices $\mathcal{I}_{\text{bdry}}^{(j)}$.
\item Remove candidates in $\mathcal{I}_{\text{bdry}}^{(j)}$: $\mathcal{C}_{\text{filtered}}^{(j+1)} = \mathcal{C}_{\text{filtered}}^{(j)} \setminus \mathcal{I}_{\text{bdry}}^{(j)}$.
\item Repeat until $\mathcal{I}_{\text{bdry}}^{(j)} = \emptyset$ or maximum 50 inner iterations reached.
\end{enumerate}
This iterative approach ensures that removing boundary points doesn't create new boundary violations in the updated Voronoi diagram.

\paragraph{Accumulation and optimization.}
We accumulate verified deep interior points across outer iterations until $|\mathcal{G}_{\text{inner}}| \geq N_{\text{inner}}$, then trim to exactly $N_{\text{inner}}$ points.
The final generator set is $\mathcal{G}_{\text{vol}} = \mathcal{G}_{\text{inner}} \cup \mathcal{G}_1 \cup \mathcal{G}_2$, where only $\mathcal{G}_{\text{inner}}$ are optimized using the CVT loss, while $\mathcal{G}_1 \cup \mathcal{G}_2$ remain fixed.
This ensures the optimized Voronoi surface is preserved exactly while optimizing volumetric mesh quality.

\section{Ablation}
\label{app:ablation}

\subsection{Noisy Point Cloud Robustness}
\label{sec:ablation-noise-robustness}

We evaluate robustness to noisy point cloud inputs by perturbing each ground-truth surface sample with zero-mean Gaussian noise of varying standard deviations. The noise magnitude is defined relative to the bounding-box diagonal of each shape to ensure scale invariance. Specifically, for a shape with bounding-box diagonal length $d$, we sample
\[
\tilde{\mathbf{x}} = \mathbf{x} + \boldsymbol{\epsilon}, 
\quad 
\boldsymbol{\epsilon} \sim \mathcal{N}\!\left(\mathbf{0}, (\sigma d)^2 \mathbf{I}\right),
\]
where $\sigma \in \{0.1\%, 0.5\%, 1.0\%, 2.0\%\}$ of the bounding-box diagonal.

\begin{figure*}[t]
    \centering
    \includegraphics[width=\linewidth]{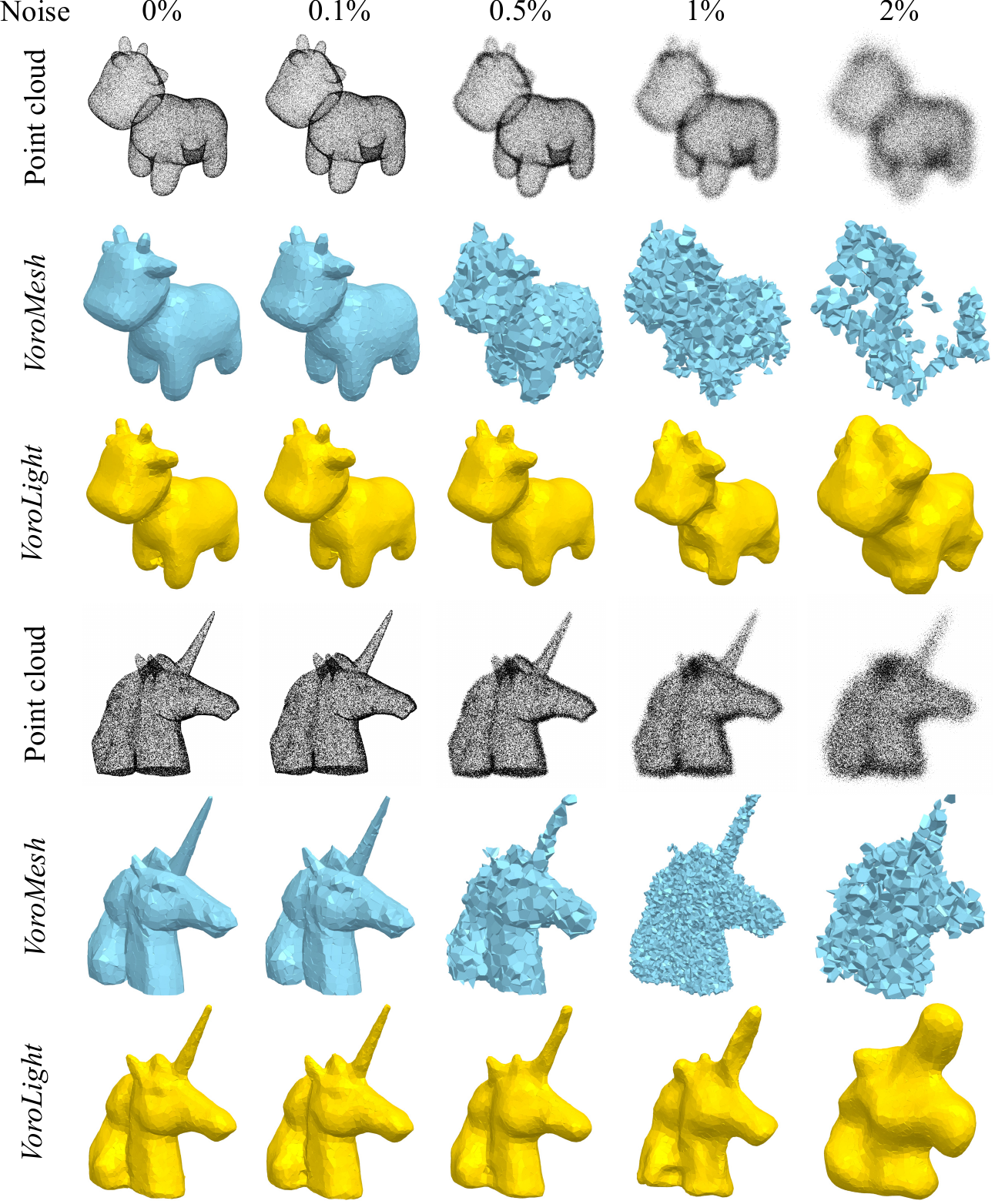}
    \caption{
Noisy point cloud reconstruction on two shapes.
We compare \textit{VoroLight} and \textit{VoroMesh} under increasing Gaussian noise levels (percentage of bounding-box diagonal).
\textit{VoroLight} results are selected from models trained with different sphere-intersection loss weights.
}
    \label{fig:noise_robustness_voromesh_comp}
\end{figure*}

We conduct experiments on two representative shapes and compare \textit{VoroLight} with \textit{VoroMesh} under increasing noise levels (Fig.~\ref{fig:noise_robustness_voromesh_comp}).

As the noise magnitude increases, both methods exhibit progressive degradation in surface quality. 
For moderate to high noise levels (0.5\%–2\%), \textit{VoroMesh} shows increasingly irregular Voronoi faces and local geometric fluctuations. 
\textit{VoroLight} also exhibits gradual smoothing of fine details at higher noise levels, while maintaining coherent global structure and watertight surfaces across all tested settings.

\begin{sidewaysfigure}
\centering
\includegraphics[width=\textheight]{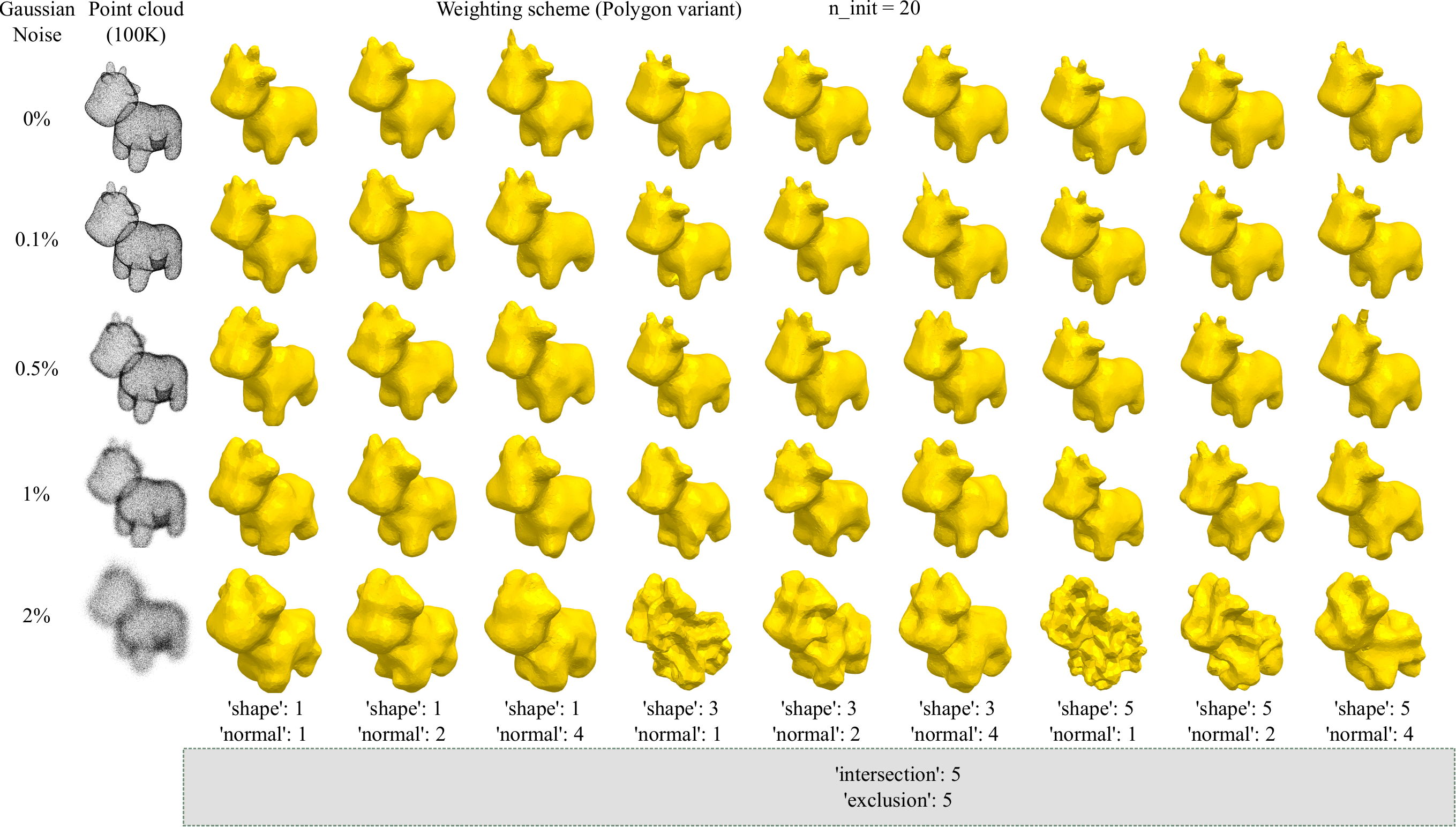}
\caption{Reconstruction quality versus input noise level for the cow shape under varying sphere-intersection loss weights in \textit{VoroLight}.}
\label{fig:noise_robustness_weights}
\end{sidewaysfigure}

\begin{sidewaysfigure}
\centering
\includegraphics[width=\textheight]{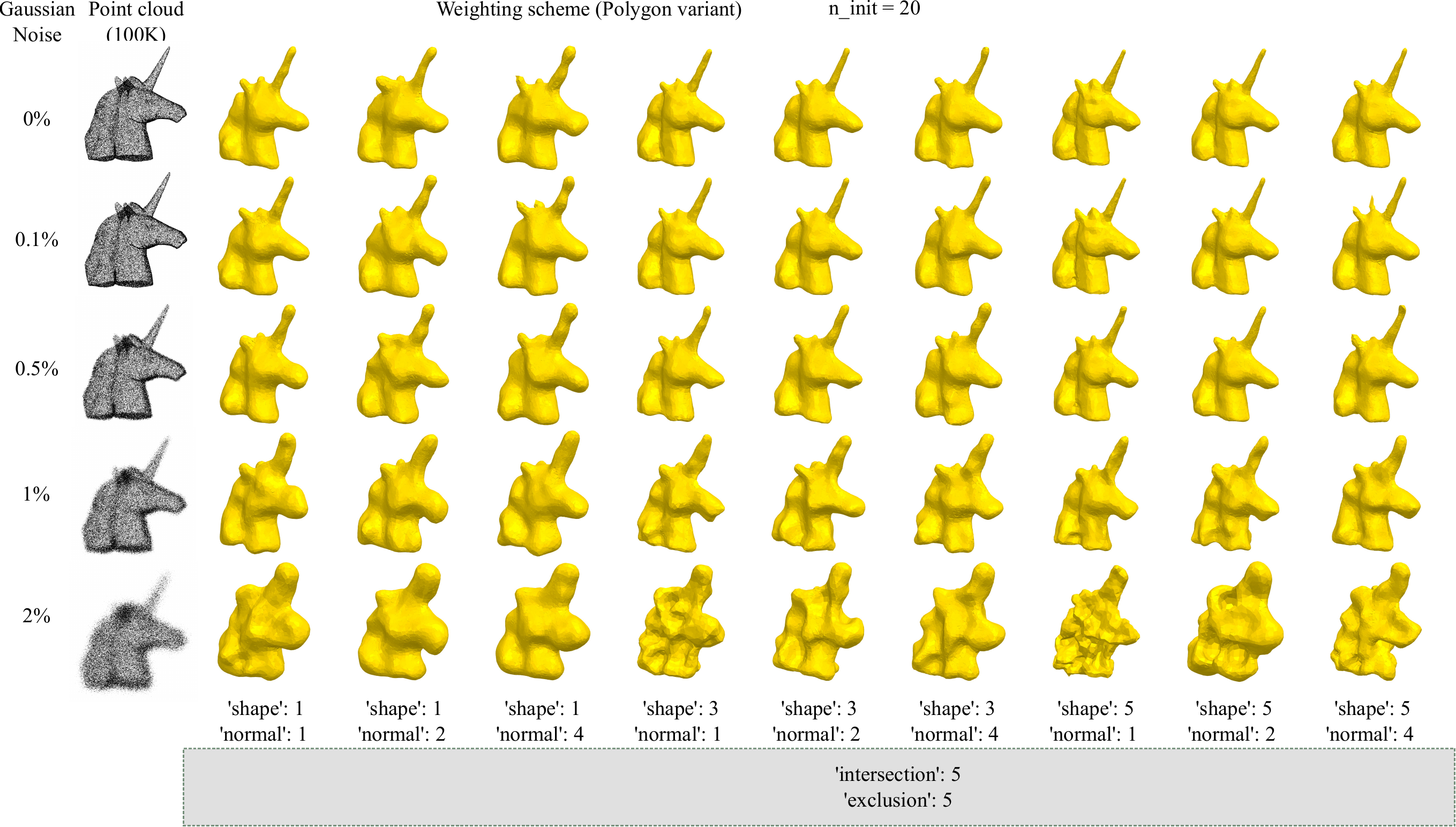}
\caption{Reconstruction quality versus input noise level for the unicorn shape under varying sphere-intersection loss weights in \textit{VoroLight}.}
\label{fig:noise_robustness_weights_B}
\end{sidewaysfigure}

We further vary the relative weights of the loss terms within the sphere-intersection objective to study their interaction with noisy supervision (Figs.~\ref{fig:noise_robustness_weights} and~\ref{fig:noise_robustness_weights_B}). 
At low noise levels, configurations with stronger shape-related supervision yield stable and well-structured surfaces with higher shape fidelity. 
As the noise level increases and the input point cloud becomes less reliable, overly strong shape supervision may introduce visible geometric distortions, whereas relatively smaller weights produce smoother reconstructions. 
Increasing the normal-consistency weight generally promotes smoother surface transitions and reduces local geometric fluctuations.
Overall, these results suggest that the sphere-intersection formulation provides a controllable geometric regularization mechanism whose optimal weighting depends on the reliability of the input point cloud.

\subsection{Initial grid resolution}
\label{sec:ablation-init-resolution}
As described in Sec.~\ref{sec:suppl_init}, $n_{\text{init}}$ denotes the number of initial grid cells along the longest side of the shape, where the initialization is a uniform cubic grid. 

\begin{figure*}[t]
    \centering
    \includegraphics[width=0.8\linewidth]{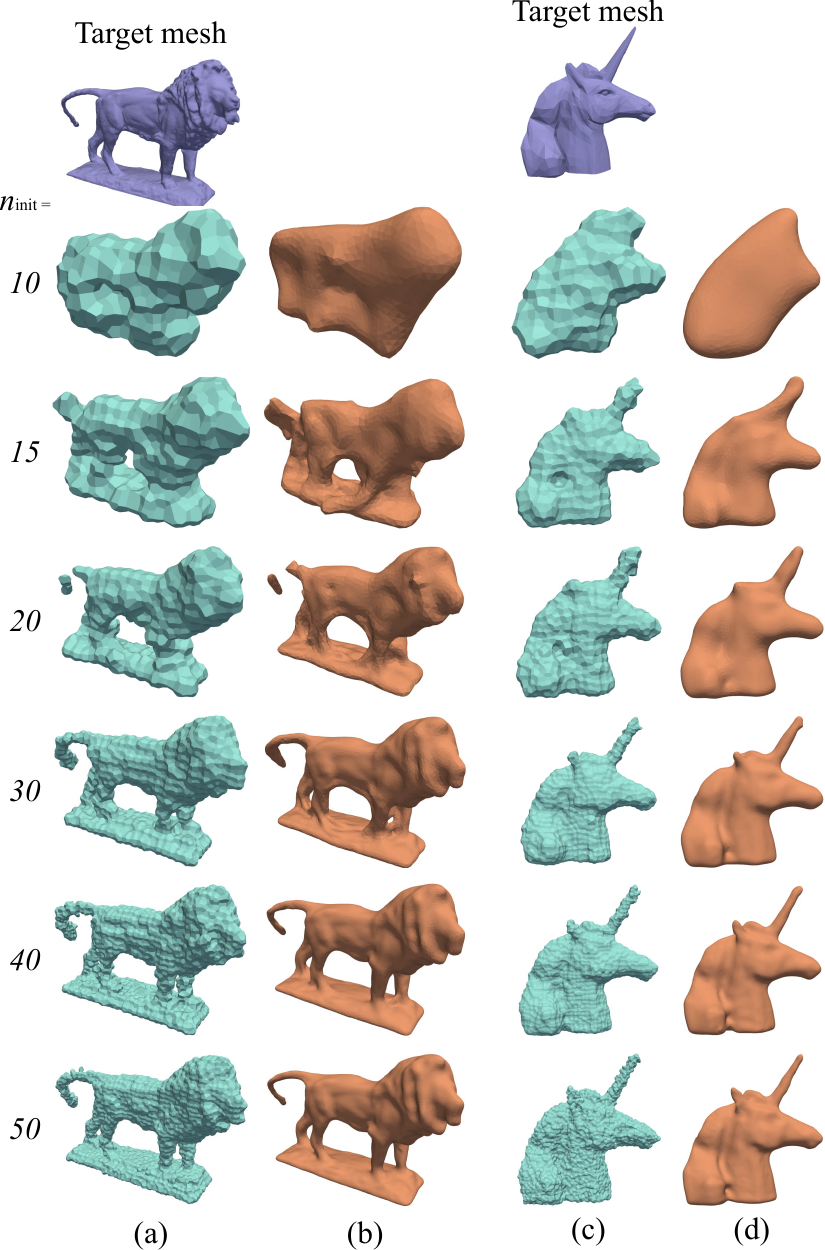}
    \caption{At different resolutions, the optimized Voronoi surface (triangle variant).
}
    \label{fig:supp_ablation_init_resolution}
\end{figure*}

Figure~\ref{fig:supp_ablation_init_resolution} shows the effect of varying this initial resolution. Across all settings, our method consistently converges to a smooth and watertight surface mesh, demonstrating strong robustness to coarse initialization. As the resolution increases, finer geometric details are recovered more faithfully. For the unicorn (left), higher-resolution initializations better capture thin structures such as the horn. For the lion model (right), the improvement is especially visible in slender regions: the tail, which may appear partially disconnected at low resolutions, becomes fully connected and anatomically coherent at higher resolutions; the legs and undercut regions start to form cleaner openings; and the curvature of the torso becomes smoother and more accurately defined.

\subsection{Volumetric mesh resolution}
\label{sec:ablation-volumetric}

\begin{figure*}[t]
    \centering
    \includegraphics[width=\linewidth]{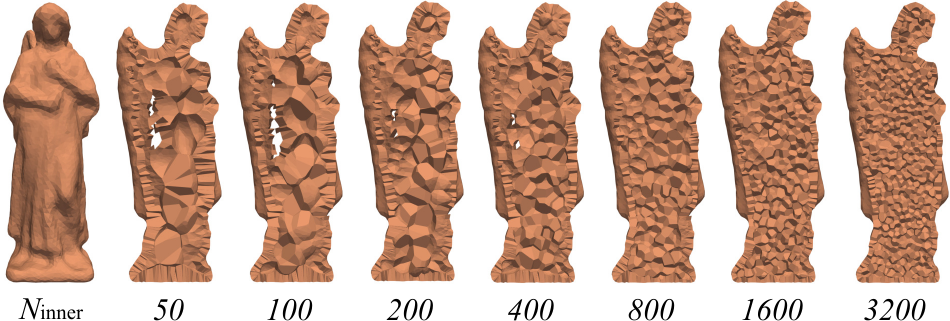}
    \caption{Ablation on the volumetric mesh resolution with varying numbers of deep inner points $N_{\text{init}}$. The surface mesh obtained uses the triangle face variant with $n_{\text{init}} = 25$.}
    \label{fig:VoroMesh_ablation_volumetric_res}
\end{figure*}

Figure~\ref{fig:VoroMesh_ablation_volumetric_res} shows volumetric meshes generated with different deep inner-point counts $N_{\text{inner}}$, following the procedure described in Sec.~\ref{sec:suppl_deep_interior_point}. The inner cells remain uniform and rounded, regularized by the CVD loss. However, because the surface cells are denser, the transition in element size from boundary cells to interior cells is not gradual. This discrepancy is reduced as the number of inner points increases and the volumetric resolution becomes closer to that of the surface. In future work, adaptive-density deep inner points could be sampled to provide smoother gradation of mesh elements---denser near the boundary and coarser in the deep interior of the shape.

\section{Additional Qualitative Results}
\label{app:additional_results}

\begin{figure*}[t]
    \centering
    \includegraphics[width=1\linewidth]{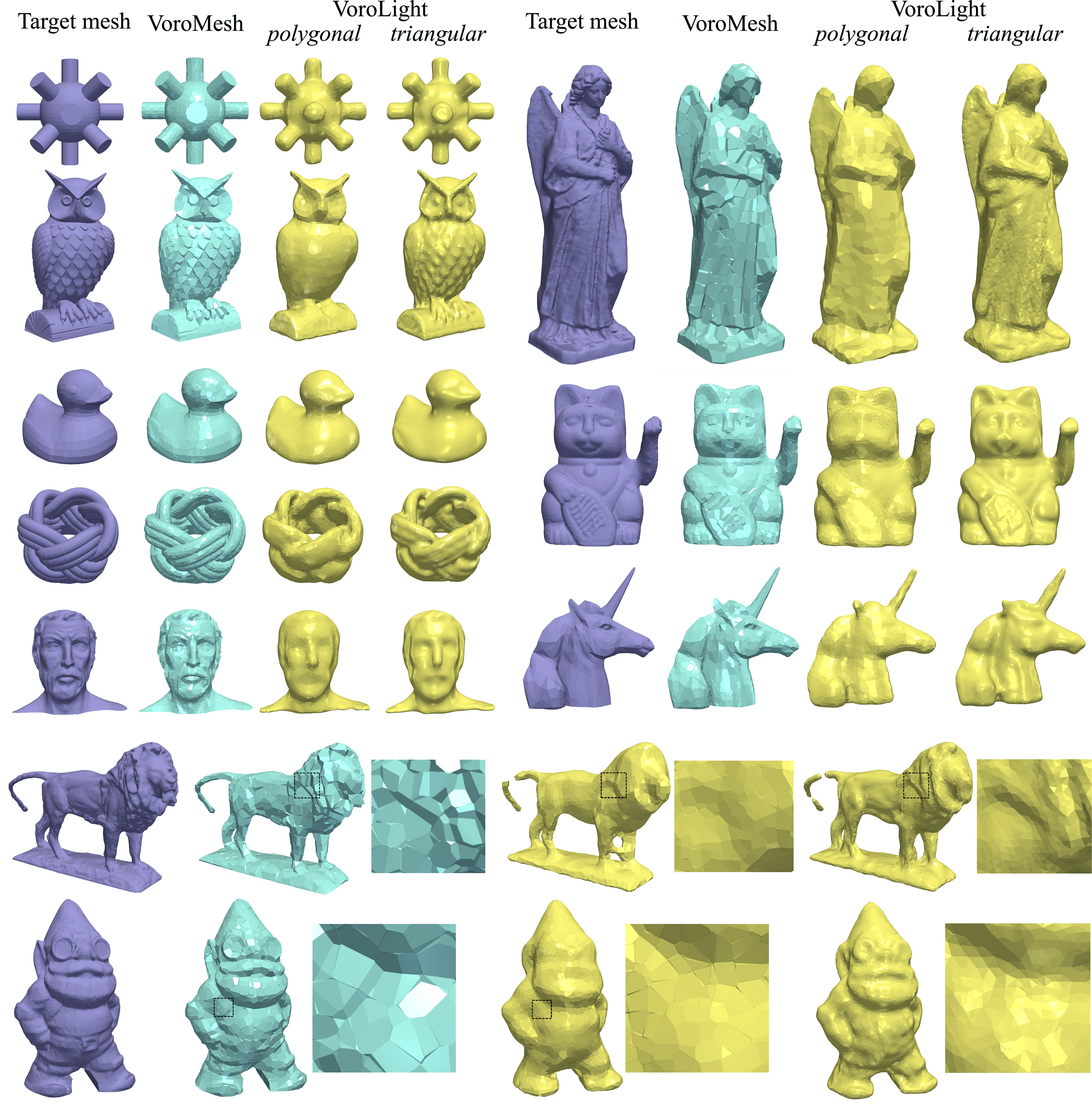}
    \caption{Comparison with \textit{VoroMesh} using mesh inputs. \textit{VoroLight} tends to smooth certain shape features but achieves near-perfect alignment of surface Voronoi faces, whereas \textit{VoroMesh} better preserves fine geometric details at the cost of slight misalignments between adjacent faces. The example shown here uses $n_{\text{init}} = 25$.}
    \label{fig:VoroMesh_Comp_more_mesh}
\end{figure*}

\begin{figure*}[t]
    \centering
    \includegraphics[width=1\linewidth]{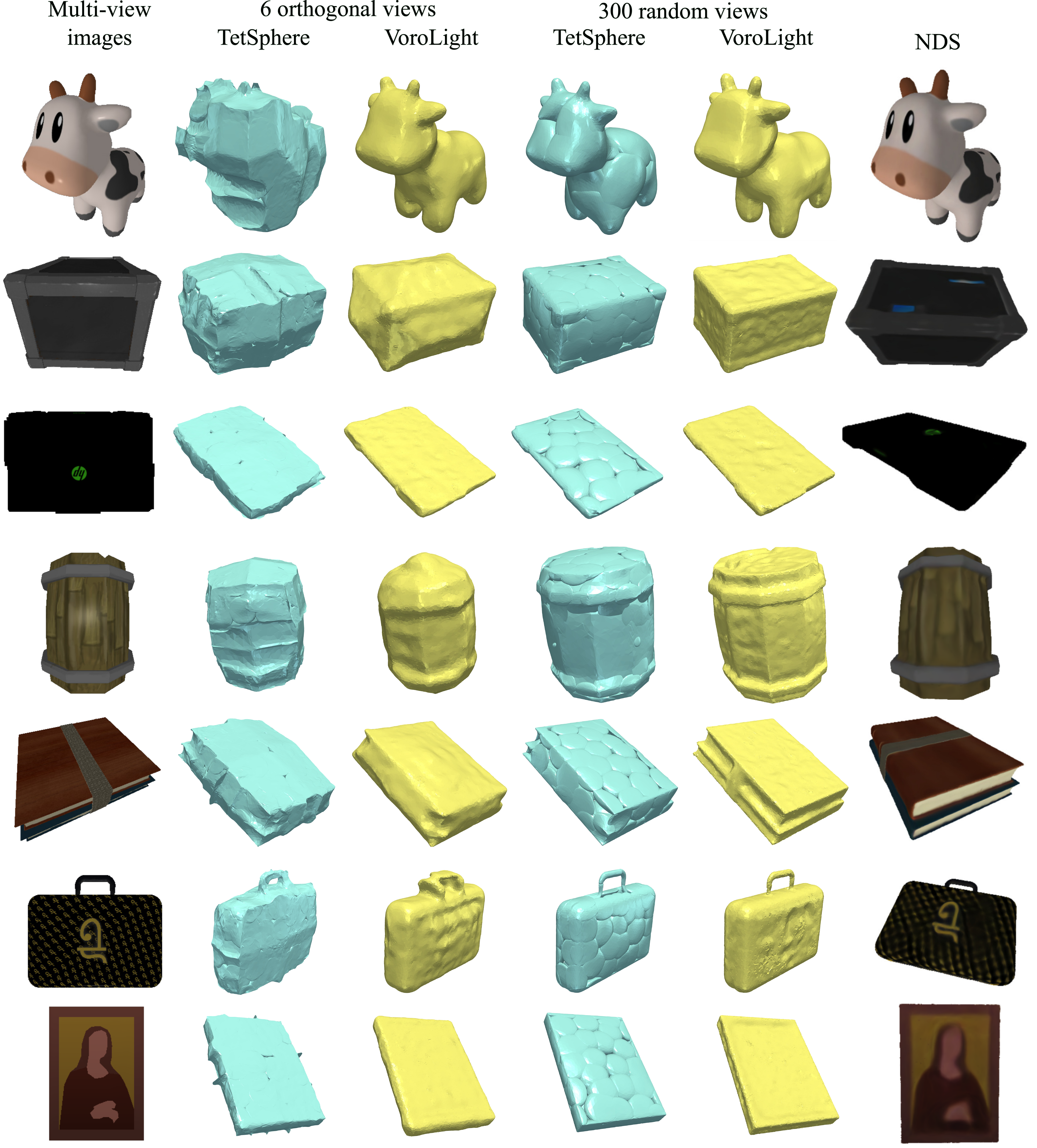}
    \caption{Comparison with \textit{TetSphere} using 6 orthogonal view and 300 random view image inputs. \textit{VoroLight} reconstructs watertight volumetric meshes with globally consistent topology, whereas \textit{TetSphere} represents shapes using locally volumetric tetrahedral meshes organized as disconnected tetrahedral clusters. The example shown here uses $n_{\text{init}} = 40$.}
    \label{fig:VoroMesh_Comp_more_300img}
\end{figure*}

\begin{figure*}[t]
    \centering
    \includegraphics[width=1\linewidth]{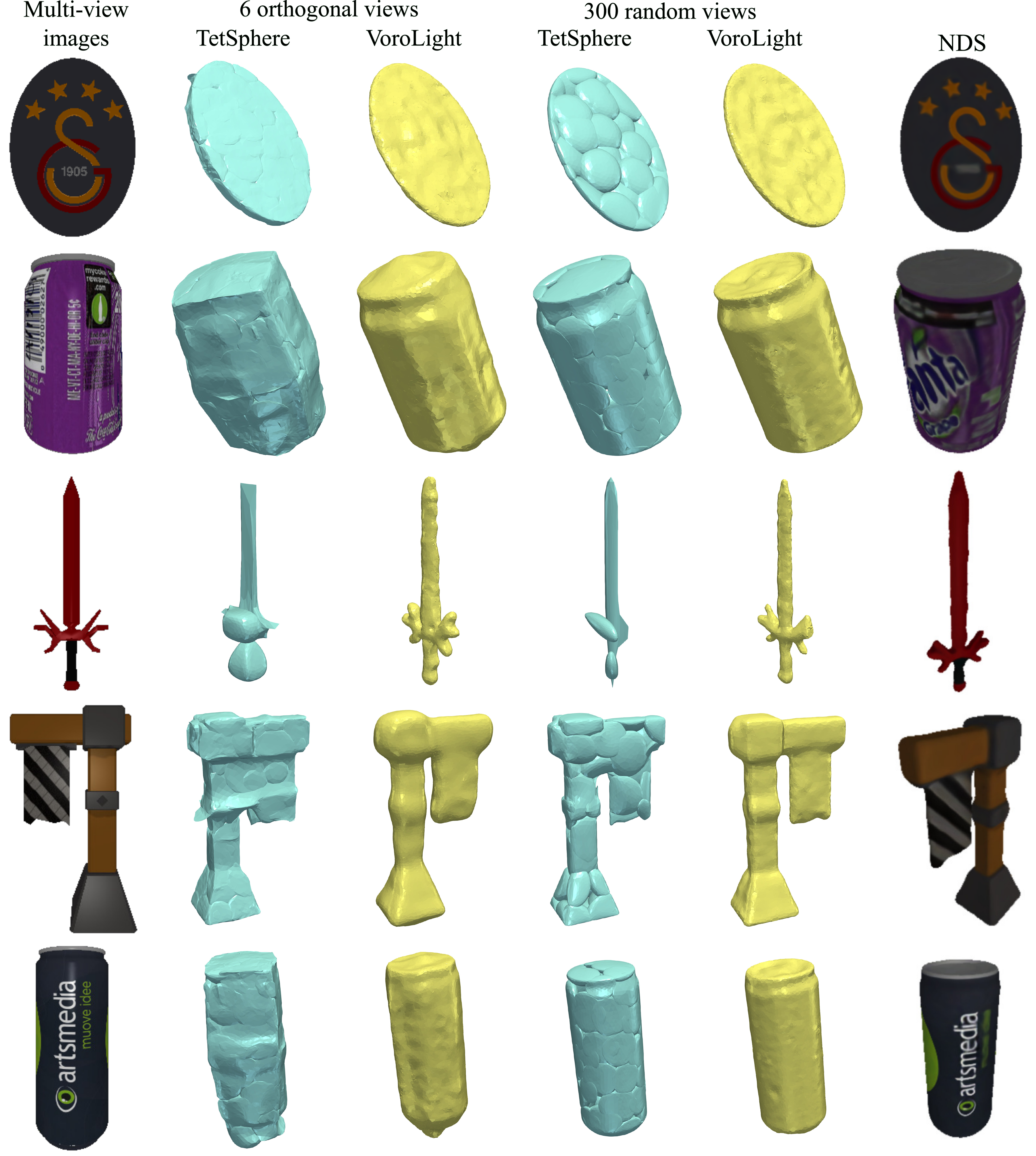}
    \caption{Comparison with \textit{TetSphere} using 6 orthogonal view and 300 random view image inputs. \textit{VoroLight} reconstructs watertight volumetric meshes with globally consistent topology, whereas \textit{TetSphere} represents shapes using locally volumetric tetrahedral meshes organized as disconnected tetrahedral clusters. The example shown here uses $n_{\text{init}} = 40$.}
    \label{fig:VoroMesh_Comp_more_300img_B}
\end{figure*}

Additional shape reconstruction results comparing \textit{VoroLight} with baseline methods.
Fig.~\ref{fig:VoroMesh_Comp_more_mesh} compares \textit{VoroLight} against \textit{VoroMesh} on mesh inputs.
Fig.~\ref{fig:VoroMesh_Comp_more_300img} and Fig.~\ref{fig:VoroMesh_Comp_more_300img_B} show comparisons with \textit{TetSphere} on 6 orthogonal view and 300 random view image inputs.

\section{Applications}
\label{sec:applications_supp}

\subsection{Single-View 3D Reconstruction}
\label{sec:app_single_image}

\begin{sidewaysfigure}
\centering
\includegraphics[width=1\linewidth]{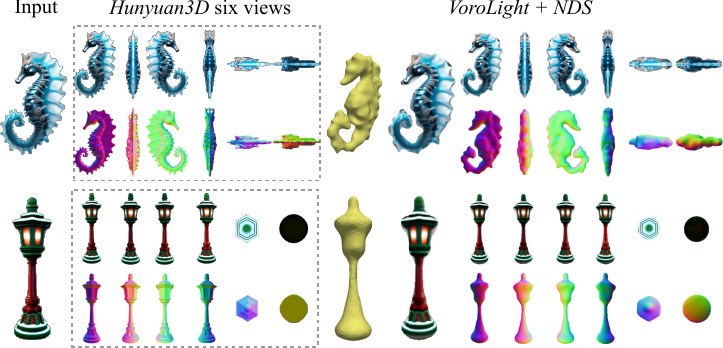}
\caption{Single-view image shape reconstruction. The example shown here uses $n_{\text{init}} = 40$.}
\label{fig:single_image_results}
\end{sidewaysfigure}

Single-view 3D reconstruction aims to infer a full, consistent 3D shape from a single RGB image, a task that is inherently ill-posed due to the loss of depth and occluded geometry.
Recent generative models, such as \textit{Hunyuan3D}~\cite{hunyuan3d22025tencent}, have demonstrated strong capabilities in synthesizing geometry-consistent multi-view images and normal maps from a single input, effectively recovering plausible 3D structure and texture.
Building on these advances, we integrate \textit{VoroLight} with \textit{Hunyuan3D} to transform such multi-view predictions into high-quality volumetric Voronoi meshes.

\begin{figure*}[t]
    \centering
    \includegraphics[width=1\linewidth]{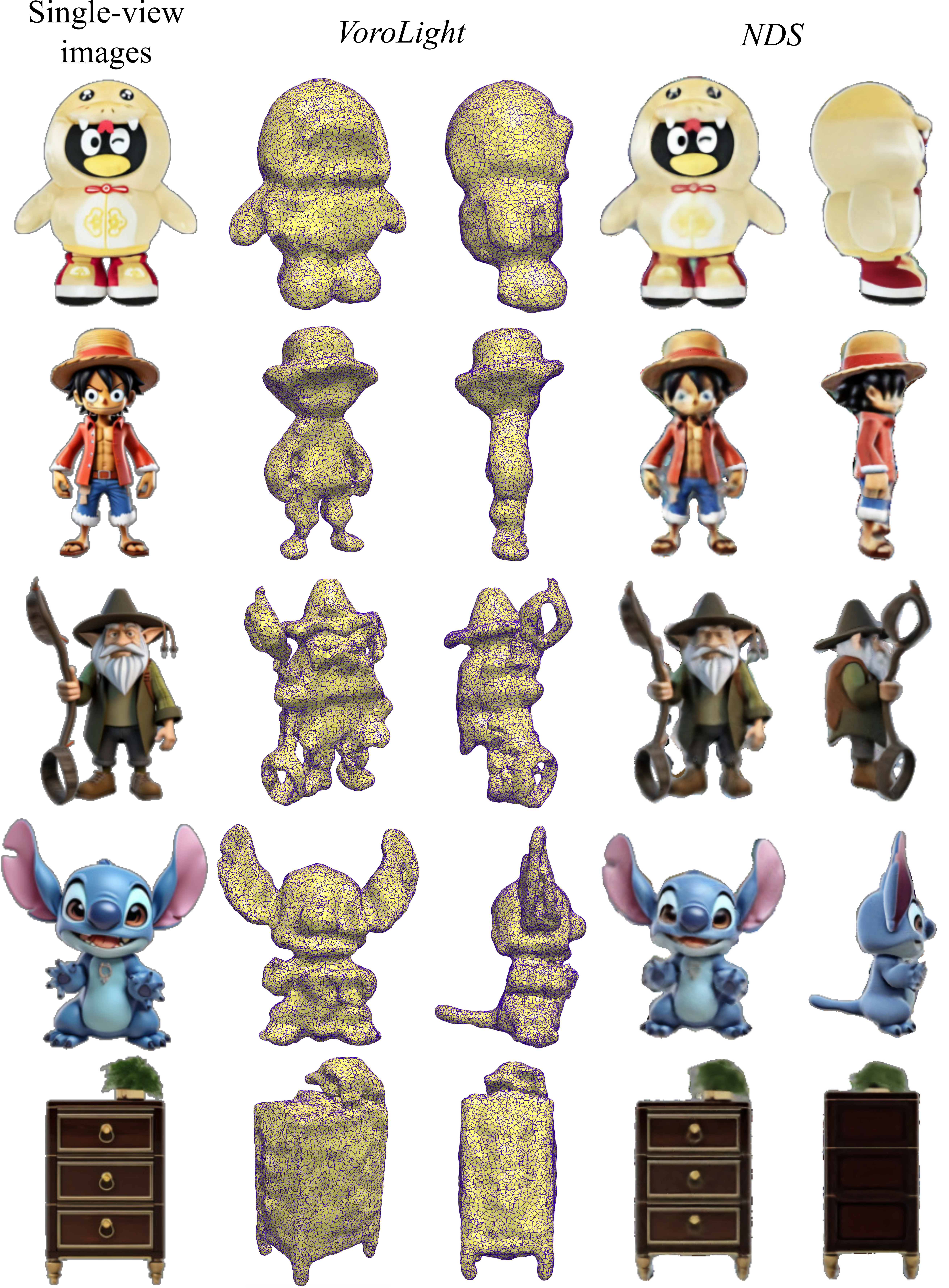}
    \caption{Additional single-view reconstruction results. We first use \textit{Hunyuan3D} to generate consistent six-view images of the target shape, and use \textit{VoroLight} to obtain the reconstructed shape. Afterwards, a neural deferred shader (\textit{NDS}) is trained to obtain the target RGB appearances. The example shown here uses $n_{\text{init}} = 40$.}
    \label{fig:VoroMesh_Comp_more_6img}
\end{figure*}

As shown in Fig.~\ref{fig:single_image_results}, from an input front-view RGB image, we use \textit{Hunyuan3D} to generate five additional RGB views (left, right, top, bottom, and back), as well as the corresponding normal fields of these total six views.
\textit{VoroLight} then reconstructs the 3D shape by optimizing the Voronoi surface to match six-view silhouette masks and normal fields via differentiable rasterization.
After obtaining this optimized geometry, the geometry is kept fixed, and a \textit{NDS} is trained on the RGB views to reproduce the predicted appearances on the mesh.
Additional single-view reconstruction results are shown in Fig.~\ref{fig:VoroMesh_Comp_more_6img}.

\subsection{Voronoi Lamp Design}
\label{sec:app_lamps}
We demonstrate polygon-face sphere training mode on artistic Voronoi lamp design (Fig.~\ref{fig:teaser_supp}): heart and sphere lamps from SDF inputs; cow lamp from single-view image; bunny lamp from surface mesh.
For 3D printing, edges are thickened into tubes that merge seamlessly at Voronoi vertices.
For artistic effect, a random subset of faces is extruded into thin solid plates while remaining faces are left hollow, with embedded cavities for LED light sources.

\end{document}